\documentclass[journal]{IEEEtran}
\usepackage[utf8]{inputenc}
\usepackage{algorithm2e}
\usepackage{mathrsfs}  
\usepackage{pbox}

\usepackage{cite}
\usepackage{amssymb}
\usepackage{diagbox}
\usepackage{url}
\usepackage{footnote}
\usepackage{hhline}
\usepackage{dsfont}
\ifCLASSINFOpdf
  \usepackage[pdftex]{graphicx}
\else
\fi
\usepackage[]{float}
\usepackage[caption = false]{subfig}

\newcommand\numberthis{\addtocounter{equation}{1}\tag{\theequation}}

\usepackage[cmex10]{amsmath}
\usepackage{amsthm}
\newtheorem{mydef}{Definition}
\newtheorem{theorem}{Theorem}
\usepackage{amsfonts}
\usepackage{color}
\usepackage[space]{grffile}
\usepackage{todonotes}
\usepackage{microtype}
\usepackage{soul}
\setstcolor{red}
\usepackage{array,multirow}
\begin{document}
%% Do not put math or special symbols in the title.
\title{Spectral Variability Aware Blind Hyperspectral Image Unmixing Based on Convex Geometry}
%\title{Can Convex Geometry-based Hyperspectral Unmixing Be Used as is to Deal with Endmember Variability?}

\author{Lucas~Drumetz,~\IEEEmembership{Member,~IEEE},  
         Jocelyn~Chanussot,~\IEEEmembership{Fellow,~IEEE}, 
         ~Christian~Jutten,~\IEEEmembership{Fellow,~IEEE},
         Wing-Kin Ma,~\IEEEmembership{Fellow,~IEEE},
         and Akira~Iwasaki
       % <-this % stops a space
        \vspace{-.2cm}
\thanks{L. Drumetz is with IMT Atlantique, Lab-STICC, UBL, Technopôle Brest-Iroise CS 83818, 29238 Brest Cedex 3, France (e-mail: lucas.drumetz@imt-atlantique.fr)}
\thanks{J. Chanussot and C. Jutten are with Univ. Grenoble Alpes, CNRS, Grenoble INP*, GIPSA-lab, 38000 Grenoble, France.  * Institute of Engineering Univ. Grenoble Alpes (e-mail: \{jocelyn.chanussot,christian.jutten\}@gipsa-lab.grenoble-inp.fr).}
\thanks{W-K. Ma is with the Chinese University of Hong Kong, Department of Electronic Engineering, Hong-Kong (e-mail: wkma@cuhk.edu.hk).}
\thanks{A. Iwasaki is with  The University of Tokyo, RCAST, Department of Advanced Interdisciplinary Studies, (e-mail: aiwasaki@sal.rcast.u-tokyo.ac.jp)}

\thanks{This work was partially funded by the Agence Nationale de la Recherche and the Direction Générale de l'Armement, by the project ANR-DGA APHYPIS, under grant ANR-16 ASTR-0027-01. L. Drumetz was also supported by a grant of the Summer Program of the Japanese Society for the Promotion of Science, JSPS-SP17206 and by a Campus France outgoing postdoctoral mobility grant, PRESTIGE-2016-4 0006.}}

\ifCLASSOPTIONcaptionsoff
  \newpage
\fi

\maketitle

\begin{abstract}
Hyperspectral image unmixing has proven to be a useful technique to interpret hyperspectral data, and is a prolific research topic in the community. Most of the approaches used to perform linear unmixing are based on convex geometry concepts, because of the strong geometrical structure of the linear mixing model. However, two main phenomena lead to question this model, namely nonlinearities and the spectral variability of the materials. Many algorithms based on convex geometry are still used when considering these two limitations of the linear model. A natural question is to wonder to what extent these concepts and tools (Intrinsic Dimensionality estimation, endmember extraction algorithms, pixel purity) can be safely used in these different scenarios. In this paper, we analyze them with a focus on endmember variability, assuming that the linear model holds. In the light of this analysis, we propose an integrated unmixing chain which tries to adress the shortcomings of the classical tools used in the linear case, based on our previously proposed extended linear mixing model. We show the interest of the proposed approach on simulated and real datasets.
\end{abstract}


% \fi
\IEEEpeerreviewmaketitle
\vspace{-0.4cm}
\section{Introduction}
\IEEEPARstart{H}{yperspectral} imaging, also known as imaging spectroscopy, is a technique which allows to acquire  information in each pixel under the form of a spectrum of reflectance or radiance values for many -- typically hundreds of -- narrow and contiguous wavelengths of the electromagnetic spectrum, usually (but not exclusively) in the visible and infra-red domains~\cite{Bioucas2013}. The fine spectral resolution of these images allows an accurate identification of the materials present in the scene, since two materials can be considered to have distinct spectral profiles. However, this identification is made harder by the relatively low spatial resolution (significantly lower than panchromatic, color or even multispectral images). Therefore, many pixels are acquired with several materials in the field of view of the sensor, and the resulting observed signature is a mixture of the contributions of these materials. Spectral Unmixing is then a source separation problem whose goal is to recover the signatures of the pure materials of the scene (called \emph{endmembers}), and to estimate their relative proportions (called \emph{fractional abundances}) in each pixel of the image~\cite{Keshava2002}.\\
In the vast majority of the studies on hyperspectral unmixing, a linear mixing model (LMM) is assumed. Each observation (pixel) is modeled as a convex combination of reference signatures, representing the pure materials of the scene. The coefficients are the fractional abundances. This model is physically valid in the so-called checkerboard configuration, i.e. when the field of view of each pixel corresponds to a flat surface, on which the materials of interest each occupy a certain area (they are mixed at a macroscopic scale)~\cite{bioucas2012,6678258}.\\
Let us denote a hyperspectral image by $\mathbf{X}\in\mathbb{R}^{L\times N}$, gathering the pixels $\mathbf{x}_{n}\in\mathbb{R}^{L}$ ($n = 1,...,N$) in its columns, where $L$ is the number of spectral bands, and $N$ is the number of pixels in the image. The signatures $\mathbf{s}_{p},\ p = 1,...,P$ of the $P$ endmembers considered for the unmixing are gathered in the columns of a matrix $\mathbf{S}\in \mathbb{R}^{L\times P}$. The abundance coefficients $a_{pn}$ for each pixel $n = 1,...,N$ and material $p = 1,...,P$ are stored in the matrix $\mathbf{A}\in\mathbb{R}^{P\times N}$. With these notations, the LMM writes:
\begin{equation}
\mathbf{x}_{n} = \sum_{p = 1}^{P} a_{pn} \mathbf{s}_{p} + \mathbf{e}_{n} = \mathbf{S} \mathbf{a}_{n} + \mathbf{e}_{n} = \mathbf{y}_{n} + \mathbf{e}_{n},
\label{LMM}
\end{equation}
where $\mathbf{e}_{k}$ is an additive noise (usually assumed to be Gaussian distributed). $\mathbf{y}_{k}$ denotes the useful signal, i.e. the noiseless data. Eq.~(\ref{LMM}) can be rewritten in a matrix form for the whole image:
\begin{equation}
\mathbf{X} = \mathbf{S}\mathbf{A} + \mathbf{E} = \mathbf{Y} + \mathbf{E},
\label{LMM_mat}
\end{equation}
where $\mathbf{E}\in \mathbb{R}^{L\times N}$ comprises all the noise values. We keep in mind the constraints on the abundances:  $a_{pn} \geq 0\ \forall (p,n)$ and $\sum_{p=1}^{P} a_{pn} = 1, \forall n$.\\
With a linear mixture, one may be tempted to resort to Independent Component Analysis approaches, which have been shown to provide excellent results in many linear source separation problems in signal processing~\cite{comon2010handbook}. However, the main assumption of this class of techniques, i.e. the independence of the sources, is violated in the case of hyperspectral unmixing, whether we consider the sources to be the endmembers, or the abundances. In the former case, the spectra of different materials of interest are typically very correlated, all the more if the materials to be unmixed are close, for instance several types of vegetation or man made materials. In the latter case, the sum-to-one constraint on the abundances immediately breaks the independence assumption.\\
That is why other methodologies were defined to tackle the unmixing problem. The most common line of attack is to rely on the strong geometrical structure provided by the LMM: the (noiseless) data is assumed to lie in a simplex whose vertices are the endmembers~\cite{Bioucas2013}. The name comes from the fact that a simplex with $P$ vertices is the simplest $(P-1)$-dimensional object that can be formed from $P$ affinely independent points embedded in a Euclidean space of dimension $L\geq P$. We give the formal definition of a simplex:
\begin{mydef}
\label{simplex}
A subset $\mathcal{S} \subseteq \mathbb{R}^{L}$ is a $(P-1)$-simplex if there exist $P$ affinely independent points $\{\mathbf{s}_{1},...,\mathbf{s}_{P}\}\in \mathbb{R}^{L}$ such that $\mathcal{S} = \emph{conv} \{\mathbf{s}_{1},...,\mathbf{s}_{P}\}$, where this denotes the convex hull of $\{\mathbf{s}_{1},...,\mathbf{s}_{P}\}$, i.e. the set 
\begin{equation}
\emph{conv} \{\mathbf{s}_{1},...,\mathbf{s}_{P}\} = \left\lbrace \mathbf{x} = \sum_{p = 1}^{P} a_{p} \mathbf{s}_{p}, \mathbf{a}\geq \mathbf{0}, \ \mathds{1}_{P}^{\top}\mathbf{a} = 1 \right\rbrace
\end{equation}    
where $\mathds{1}_{\cdot}$ denotes a vector of ones whose size is given in index.
\end{mydef}

In other words, a $(P-1)$-simplex is the convex hull of $P$ affinely independent points in $\mathbb{R}^{L}$.
\begin{figure}
\centering
\includegraphics[scale=0.15]{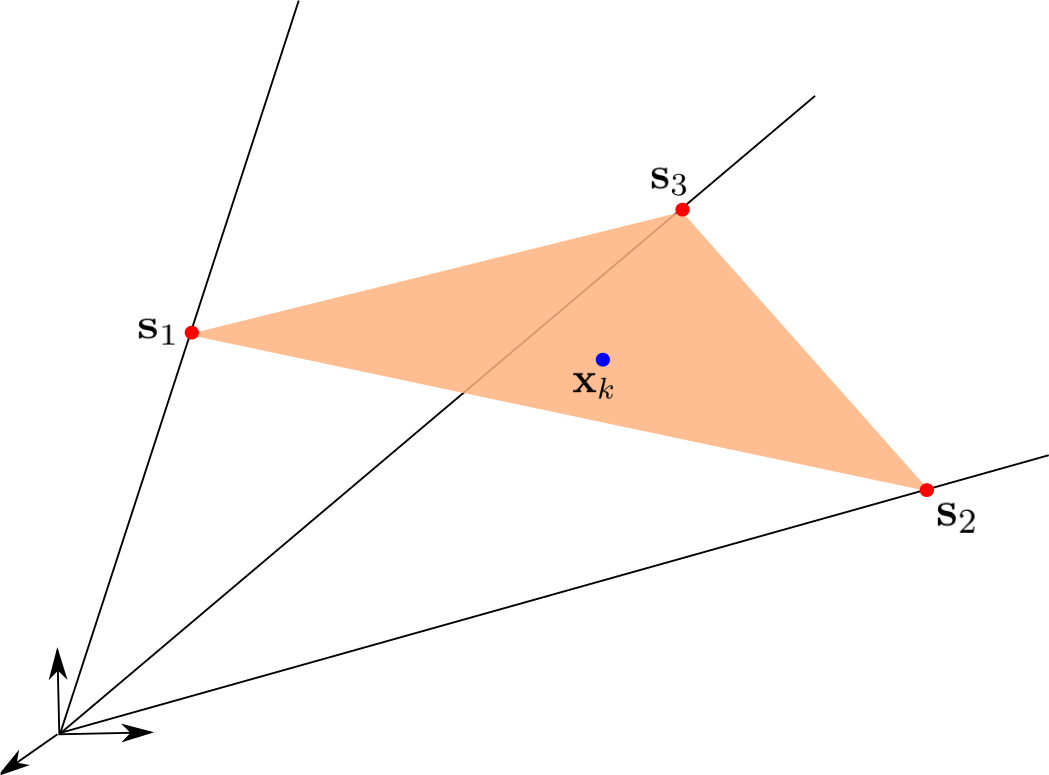}
\caption{Geometric interpretation of the LMM in the case of three endmembers (red dots) in a three-dimensional embedding space. The axes represent a basis of the linear subspace spanned by the endmembers.}
\label{simplex_geom}
\end{figure}
An illustration of the geometrical interpretation of the LMM is given in Figure~\ref{simplex_geom}. From this observation, the typical unmixing processing chain is usually divided into three steps, which all rely on the convex geometry of the problem:
\begin{enumerate}
\item Estimating the number of endmembers to consider. This is a very hard and ill-posed problem in itself (because there is no such thing as an optimal number of endmembers in real data, among other reasons) and many algorithms have been considered in the community to try to obtain a good estimate~\cite{Robin2015}. The so-called Intrinsic Dimensionality (ID) of the data if often used as an estimate of the number of endmembers. The usual definition of the ID is that it is the dimension of the signal subspace, i.e. the dimension of the column space of the (noiseless) data matrix:
\begin{mydef}
\label{ID}
The Intrinsic Dimension (ID) of a dataset, $\mathbf{x}_1,\ldots,\mathbf{x}_n$, is the dimension, $d$, of the vector subspace spanned by the signals, $\mathbf{y}_1,\ldots,\mathbf{y}_n$.
\end{mydef}
In the noiseless linear case, if the LMM is valid,  $d = \textrm{span}(\mathbf{SA})$, and this indeed corresponds to the number of endmembers (if $\mathbf{S}$ and $\mathbf{A}$ have full column and row ranks, respectively, which is a quite reasonable assumption). One of the best-known algorithms to perform this estimation is the Hyperspectral Subspace Identification by Minimum Error (HySIME)~\cite{Bioucas2008}.
\item Extracting the spectra corresponding to the endmembers, a procedure referred to as endmember extraction. Then again, many Endmember Extraction Algorithms exist in the literature to tackle this problem, with various assumptions, the main one being the presence in the data of pure pixels, i.e. pixels in which only one material of interest is present~\cite{Plaza2004}. These algorithms try to exploit the geometry of the problem by looking for extreme pixels in the data, which are the endmembers if the LMM holds. A popular endmember extraction algorithm using the pure pixel assumption is the Vertex Component Analysis (VCA)~\cite{Nascimento2005}.
\item Finally, estimating the abundances using the data and the extracted endmembers. This step is usually carried out by solving a constrained optimization problem:
\begin{equation}
\underset{\mathbf{A}\geq \mathbf{0}, \ \mathds{1}_{P}^{\top}\mathbf{A} = \mathds{1}_{N}^{\top}}{\textrm{arg min}} \ ||\mathbf{X}-\mathbf{SA}||_{F}^{2},
\label{FCLSU}
\end{equation}
where $||\cdot||_{F}$ denotes the Frobenius norm. Solving this problem is often referred to as Fully Constrained Least Squares Unmixing (FCLSU)~\cite{Heinz2001}. Other methods based on Nonnegative Matrix Factorization (NMF) are able to jointly compute the abundances and refine the endmember estimation at the same time~\cite{lee1999learning}.
\end{enumerate}

\begin{figure}
\centering
\begin{minipage}{0.45\linewidth}
\includegraphics[scale=0.1]{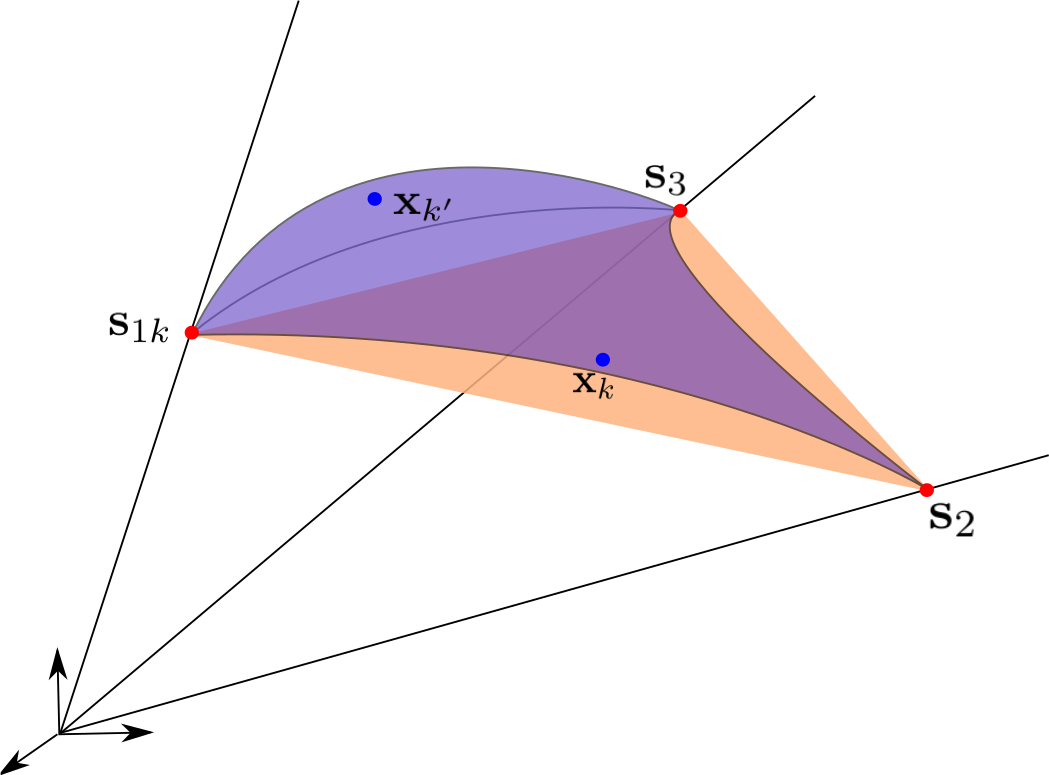}
  \centerline{(a)}\medskip
\end{minipage}
\begin{minipage}{0.45\linewidth}
\includegraphics[scale=0.1]{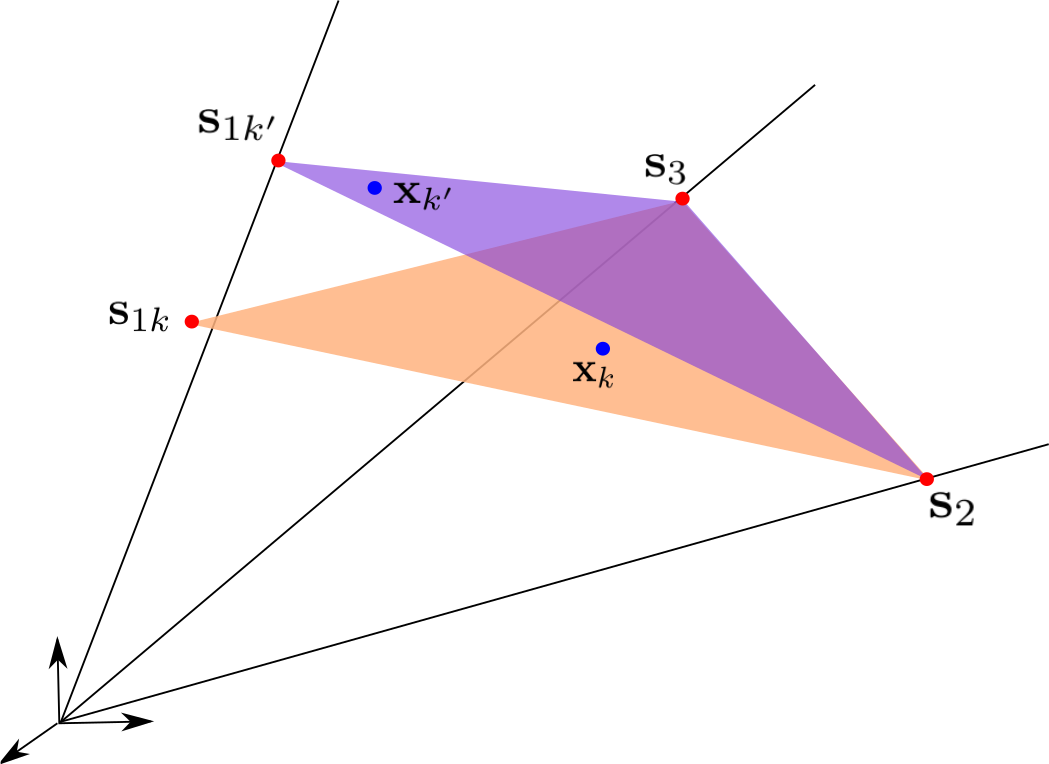}
  \centerline{(b)}\medskip
\end{minipage}
\caption{Geometric interpretation of (a) a nonlinear mixing model (in purple) and (b) of endmember variability in a LMM framework. The new observation $\mathbf{x}_{k'}$ is not in the original simplex, but can be explained by changing the model.}
\label{var}
\end{figure}

Over the years, several limitations of the LMM became apparent, the main ones being identified as nonlinearities in the mixing process, the intra-class variability of each material, and the dependence on pure pixels.
\begin{itemize}
\item Nonlinearities occur when the mixture of the materials takes place at an intimate level, e.g. in particulate media such as sand~\cite{Hapke2012}, or when the light undergoes multiple reflections before reaching the sensor, which can happen in tree canopies or urban scenarios~\cite{Heylen2014,6678284}. An example of the geometry of a nonlinear mixing model is shown in Fig.~\ref{var}~(a).
\item Considering endmember variability, on the other hand, simply means that we cannot reasonably assume that a single spectrum can fully represent a material in all its diversity. Several factors can indeed change the signature of a material, be it due to changing illumination conditions and topography, which locally change the geometry of the hyperspectral acquisitions, or the intrinsic variability of the materials, for instance the effect of a change in chlorophyll concentration in green vegetation~\cite{zare2014,drumetz2016endmember}. A geometrical interpretation of endmember variability is shown in Fig.~\ref{var}~(b).
\item When endmember extraction algorithms are used, one last issue is the necessity of pure pixels. Pure pixels are pixels whose abundance is one for one the materials and zero for the rest. Their presence is mandatory for geometry-based endmember extraction algorithms to work. Various methods have been designed to unmix data which do not satisfy this assumption~\cite{berman2004ice,Li2008,chan2009convex}. They are based on minimizing the volume of the simplex used in the unmixing, so as to enclose the data in a simplex whose vertices are going to be the endmembers.
\end{itemize}

In this context, it is natural to wonder to what extent the linear unmixing chain still applies to the problem, and how each of the steps transfers to more complex mixing models. This question was never really adressed by the community, even though some statements about the robustess of certain algorithms to some of the aforementioned phenomena can be found here and there in the literature. For instance, concerning ID estimation, authors in~\cite{Bioucas2008} state that nonlinear models are still embedded in a linear subspace of dimension much lower than the number of spectral bands considered, and as such ID can still provide an upper bound for the number of endmembers. More information on the geometry of popular nonlinear models can be found in the recent overviews~\cite{Heylen2014,6678284}. In~\cite{6678284}, endmember extraction algorithms are considered to be able to correctly identify endmembers in mildly nonlinearly mixed data because for many models, endmembers are still extreme points of the nonlinear data manifold. However, for certain models, the nonlinear manifolds have extremities which are not the linear endmembers anymore~\cite{Heylen2014}. The pure pixel assumption has been the subject of several theoretical studies to determine in which configurations the endmembers could be identified in spite of the absence of pure pixels, and theoretical results on their efficiency can be derived, but then again, they are based on the LMM~\cite{lin2015identifiability}.\\
In this paper, we propose to discuss the validity of all these convex geometry-based techniques, in a context where the linear mixing model is still valid, but considering endmember variability. The reason for this choice is twofold: we consider that endmember variability generally affects the observations more than nonlinearities, which are predominant only in very specific cases, some of which are mentioned above. Second, with an appropriate modeling of endmember variability, the problem, although strictly speaking nonlinear, retains a strong geometrical structure, and thus theoretical results as well as geometrical concepts still apply with some modifications. In the nonlinear case, we will simply state that geometrical endmember extraction algorithms are still useful when the nonlinear model preserves the extreme nature of endmembers in the dataset, and ID estimation can still provide an upper bound of the number of endmembers to use. The efficiency of volume regularization is a more complex topic, but since the data manifold is no longer a simplex, it is likely that these methods will fail in strongly nonlinear scenarios. Even though theoretically powerful and appealing, combining nonlinear effects and endmember variability leads to very complex models whose efficiency in practice remain to be proven~\cite{revel2016linear,halimi2016hyperspectral}. Besides, the extended mixing model we will use throughout this paper to model brightness variations has recently been shown to be able to estimate the abundances of nonlinear mixing models modeling multiple interations to some extent, both theoretically and experimentally~\cite{drumetz2017relationships}. For all these reasons, we will consider nonlinear models out of the scope of the present paper.\\
Our contributions are as follows: we propose an analysis of the convex geometry based concepts used in hyperspectral image unmixing in the context of illumination induced, as well as intrinsic variability (with a precise perimeter for both these terms), in a LMM framework. We discuss ID estimation, endmember extraction, as well as pixel purity. Besides, with the insight gained from this analysis, we propose an integrated unmixing chain for this context, which is still based on convex geometry concepts, but adapts them to the new geometry induced by the presence of variable endmembers. We describe the proposed method (first outlined in~\cite{drumetz2018endmembers}), and show its relevance on several datasets, both semi-synthetic and real, with quantitative and qualitative analyses of the results.\\
The remainder of this paper is organized as follows: Section~\ref{ev} will review how spectral variability can be taken into account in the unmixing problem, and precise how we will consider it in this paper. Based on previous studies, Section~\ref{analysis} will provide an analysis of the behavior of different convex geometry concepts used in classical linear unmixing when endmember variability is considered. We will suggest some leads to circumvent the limitations of current approaches, which we will convert into an adapted unmixing chain and algorithm, as presented and described in Section~\ref{chain}. The results of the proposed approach on a synthetic and a real dataset are discussed in Section~\ref{results}. We gather some concluding remarks in Section~\ref{conclusion}.
\section{Endmember Variability}
\label{ev}
Endmember variability simply refers to the fact that one material cannot be completely represented by a single spectral signature, since many factors can induce modifications on the observed spectra corresponding to one material. The two main factors we consider here are the variations induced by changing illumination conditions (``extrinsic" variability) and all the modifications induced by changes in the physico-chemical composition of the materials (intrinsic variability). Before describing these two different types of variability and some of the existing approaches to tackle them, we formalize mathematically the notion of spectral variability in the unmixing problem, by restating the LMM as a \emph{space varying} linear model. In this case, endmember variability essentially amounts to allow the endmember matrix to vary from one pixel to the other, within a linear model~\cite{drumetz2016endmember}:
\begin{equation}
\mathbf{x}_{n} = \mathbf{S}_{n} \mathbf{a}_{n} + \mathbf{e}_{n}
\label{LMM_var}
\end{equation}
Of course, without further modeling on $\mathbf{S}_{n}, \forall n$, this model is very general and solving the inverse problem of recovering endmembers and abundances is extremely ill-posed. We will not provide a full catalog of all the existing models and algorithms designed to tackle endmember variability, and refer to the recent reviews for the interested reader~\cite{Zare2013,drumetz2016endmember}. Most of these models require to obtain reference endmembers from which the variability will be extrapolated. These are usually obtained using endmember extraction algorithms (designed for purely linear models). We will come back to this issue later in this paper.
\subsection{Illumination-induced variability}
\begin{figure}
\begin{center}
\includegraphics[scale=0.35]{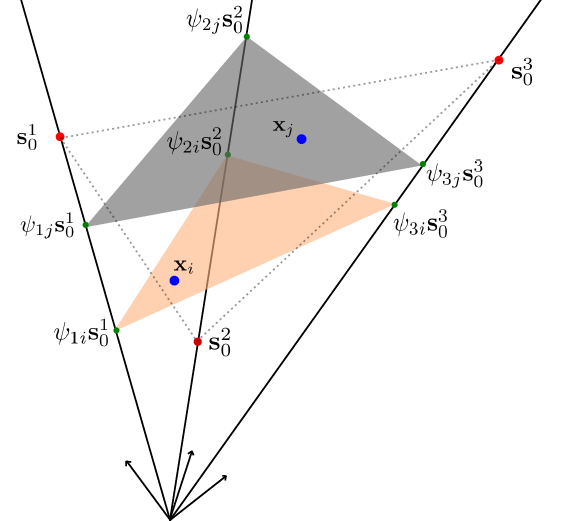}
\caption{Geometric interpretation of the ELMM in the case of three endmembers. In blue are two data points, in red are the reference endmembers and in green are the scaled versions for the two considered pixels. The simplex used in the LMM is shown in dashed lines.}
\label{ELMM_draw}
\end{center}
\end{figure}

Changing illumination conditions can have a tremendous impact on the observed spectral signature of the materials, regardless of changes in their composition. Simple examples include shadowed materials, whose signatures are lower than if they were receiving full illumination from the sun. Besides, reflectance and radiance, the physical quantities used in hyperspectral imaging, are both dependent on the geometry of the acquisition, i.e. the incidence angle of the light with the material, and the viewing angle of the sensor. These are not fixed on all the support of a given image, since they are dependent on the topography of the scene, which locally changes the geometry of the acquisition. Complex radiative transfer models were designed to describe these physical phenomena, one of the most famous being the model derived by Hapke~\cite{Hapke2012}. However, this model is much too complex to be directly used in blind unmixing, and also depends on many empirical parameters (the albedo of the materials, the acquisition angles, photometric parameters of each material), which are rarely (if at all) available in real scenarios.\\
The application of Hapke model to generate variants of a given spectrum was theoretically and experimentally shown, however, to be reasonably approximated by (nonnegative) scaling variations of this signature~\cite{drumetz2019spectral,Nascimento2005-ICA, drumetztip,hong2019augmented}, i.e. we can reasonably model $\mathbf{S}_{n} = \psi_{n} \mathbf{S}_{0}$, where $\psi_{n}$ is a scaling factor accounting for brightness variations of a reference endmember matrix $\mathbf{S}_{0}$. Eq.~\eqref{LMM_var} then becomes:
\begin{equation}
\mathbf{x}_{n} = \psi_{n} \sum_{p = 1}^{P} a_{pn} \mathbf{s}_{0p} + \mathbf{e}_{n}.
\label{LMM_var_mat}
\end{equation}
Note that with this model, the product $\phi_{pn} \triangleq \psi_{n} a_{pn}$ is no longer required to sum to one, which means in practice it can be simply estimated by nonnegative least squares (i.e. by dropping the usual sum-to-one constraint). Then, to split the product into abundances and scaling factors, one can simply sum these quantities over the materials, for a given pixel~\cite{Veganzones2014_ELMM}:
\begin{equation}
\sum_{p=1}^{P} \phi_{pn} = \sum_{p=1}^{P} \psi_{n} a_{pn} = \psi_{n} \sum_{p=1}^{P}  a_{pn} = \psi_{n},
\end{equation}
if we reintroduce the sum-to-one constraint on the \emph{actual} abundances. Then the abundances can be reestimated by scaling the product $\phi_{pn}$, as $a_{pn} = \frac{\phi_{pn}}{\psi_{n}}$. $\psi_{n}$ is never zero since at least one material is present in each pixel. This technique will be referred to as Scaled (partially) Constrained Least Squares Umixing (SCLSU) in the remainder of the paper.\\
In order to be able to explain material specific (i.e. photometry-related) effects on the spectra, the scaling factors can be further assumed to depend on the considered endmember, giving the so-called (full) Extended Linear Mixing Model (ELMM)~\cite{drumetztip}:
\begin{equation}
\mathbf{x}_{n} = \sum_{p = 1}^{P}\psi_{pn}  a_{pn} \mathbf{s}_{0p} + \mathbf{e}_{n}.
\label{ELMM}
\end{equation}
In this case, we have $\mathbf{S}_{n} = \mathbf{S}_{0} \boldsymbol{\psi}_{n}$, where $\boldsymbol{\psi}_{n}\in \mathbb{R}^{P\times P}$ is a diagonal matrix incorporating the scaling factors for each material on its diagonal.\\
The two scaling factor models can also be expressed globally in the whole image, connecting them to classical NMF models:
\begin{equation}
\mathbf{X} = \mathbf{S}_{0}(\mathbf{A}\odot \boldsymbol{\Psi}) + \mathbf{E} = \mathbf{S}_{0}\boldsymbol{\Phi} + \mathbf{E}
\end{equation}
where $\odot$ is the Schur-Hadamard (elementwise) product, $\boldsymbol{\Psi}\in \mathbb{R}^{P\times N}$ gathers all the scaling factors, and $\boldsymbol{\Phi} \triangleq  \mathbf{A} \odot \boldsymbol{\Psi}$. This formulation also reveals there is an inherent multiplicative ambiguity between abundances and scaling factors, and hence further assumptions are needed to split $\boldsymbol{\Phi}$ into two terms. SCLSU asssumes the scaling factor in a pixel to be the same for all endmembers, and the algorithm used in~\cite{drumetztip} for the full ELMM makes additional statistical assumptions.\\
Nevertheless, the model possesses a nice geometrical interpretation (shown in Fig.~\ref{ELMM_draw}), generalizing the LMM: endmembers are no longer constrained to be single points, but can lie anywhere on lines joining the origin and the reference endmembers (the columns of $\mathbf{S}_{0}$). In both cases, the data are no longer constrained to lie in a simplex, but in a convex cone spanned by the reference endmembers (or any scaled version of them, for that matter). More precisely, the data lie in a polyhedral cone: 
\begin{mydef}
\label{cone}
A subset $\mathcal{C} \subseteq \mathbb{R}^{L}$ is a polyhedral cone (or a finitely generated convex cone) with $P$ generators if there exist $P$ linearly independent points $\{\mathbf{s}_{1},...,\mathbf{s}_{P}\}\in \mathbb{R}^{L}$ such that $\mathcal{C} = \emph{cone} \{\mathbf{s}_{1},...,\mathbf{s}_{P}\}$, where this denotes the conical hull of $\{\mathbf{s}_{1},...,\mathbf{s}_{P}\}$, i.e. the set 
\begin{equation}
\emph{cone} \{\mathbf{s}_{1},...,\mathbf{s}_{P}\} = \left\lbrace \mathbf{x} = \sum_{p = 1}^{P} \phi_{p} \mathbf{s}_{p}, \boldsymbol{\phi}\geq \mathbf{0}\right\rbrace
\end{equation}    
\end{mydef}
On the edges of this cone, the scaling factors encode the position of the local endmembers w.r.t. the references.
\subsection{Intrinsic variability}
\begin{figure}
\begin{minipage}{0.49\linewidth}
\centering
\includegraphics[scale=0.12]{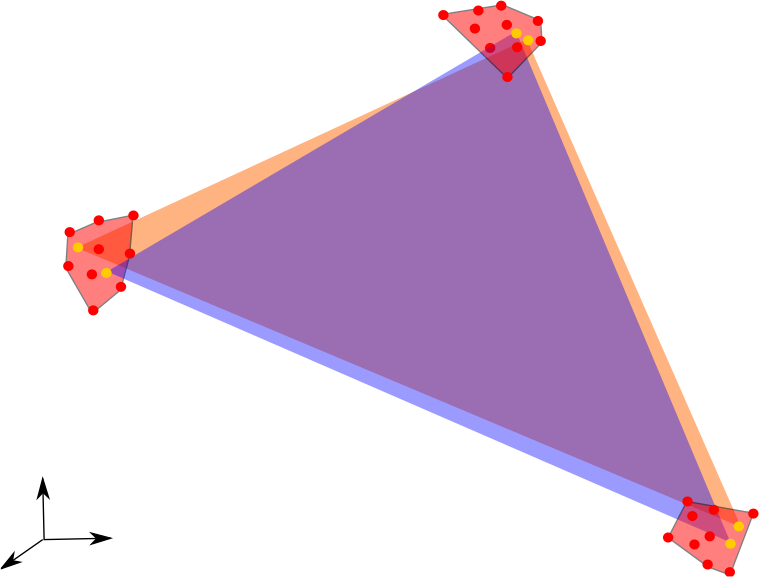}
\centerline{(a)}\medskip
\end{minipage}
\begin{minipage}{0.49\linewidth}
\centering
\includegraphics[scale=0.12]{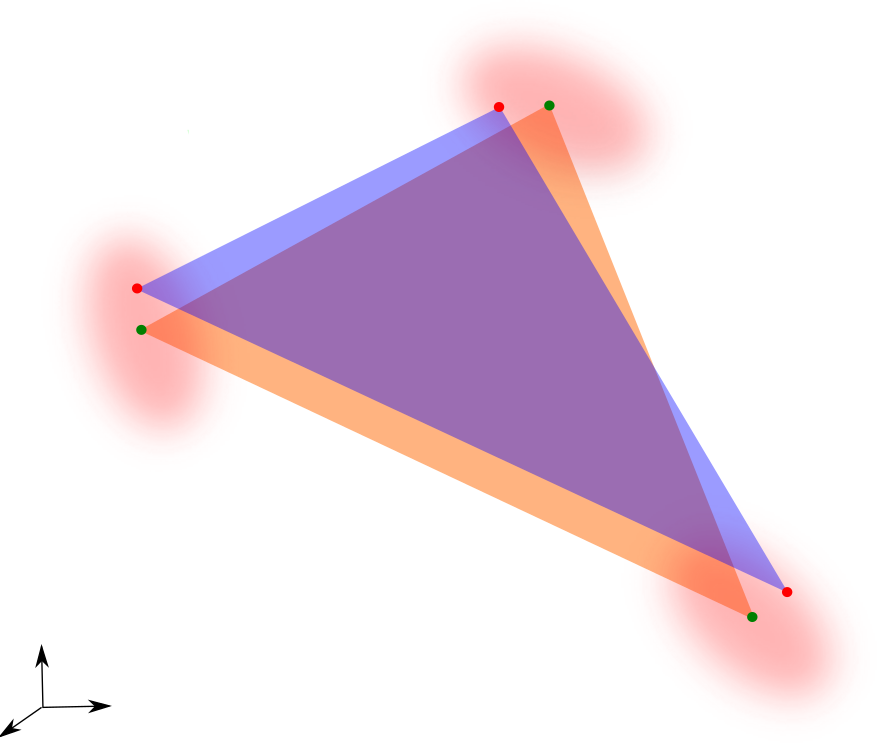}
\centerline{(b)}\medskip
\end{minipage}
\caption{Geometric interpretation of (a) using FCLSU on the whole extracted dictionary. The red polytopes are the convex hull of the different bundles. The yellow points are accessible endmembers when using FCLSU, whereas they were not extracted by the EEA (b) Purely statistical models, where endmembers are drawn from statistical distributions.}
\label{bundles_fclsu}
\end{figure}

If illumination related variability can be physically modeled in a tractable way for hyperspectral image unmixing, the situation is much more complex for the intrinsic variability of the materials, due to its material specific nature and to the numerous factors which should be taken into account for each different material. To cope with this inherent hurdle for the unmixing application, it hardly comes at a surprise that researchers have turned to statistical models to capture this phenomenon instead. One can categorize statistical models into two broad classes: bundle-based methods, and model-based methods. We briefly summarize some of these models below.\\
The concept of spectral bundles was introduced in~\cite{somers2012}, under the name ``Automated Endmember Bundles" (AEB). The underlying idea is to represent endmember variability as a set of candidate signatures representing different instances of each material, and include each of them in the model as potential endmembers. In the context of \emph{blind} unmixing, these candidates have to be extracted from the image.\\
To do that, several subsets of the image are randomly selected (possibly sampling without replacement to ensure that different endmember instances are selected every time). An endmember extraction algorithm is run on each of these subsets, to extract as many signatures as the number of endmembers considered globally. If there is at least one pure pixel in each subset for each material, then different instances of each endmember are likely to be selected. All the candidate endmembers are then gathered in a dictionary of candidate endmembers. However, since most endmember extraction algorithms are stochastic, the extracted sources are not \emph{aligned}, i.e. the order of the endmembers is not the same from one subset to the other, and there is a priori no grouping of the different signatures into classes containing different instances of the same endmembers. To solve this problem, a clustering step is required, in order to group the signatures into $P$ bundles of candidate endmembers for the different materials. This can be done for instance with the k-means algorithm, usually using the Spectral Angle Mapper (SAM)
\begin{equation}
SAM(\mathbf{s}_1,\mathbf{s}_2) = \textrm{acos}\left(\frac{\mathbf{s}_{1}^{\top}{\mathbf{s}_{2}}}{||\mathbf{s}_{1}||_{2}||\mathbf{s}_{2}||_{2}} \right)
\label{SAM}
\end{equation}
as the similarity measure, since it is known to be insensitive to scaling variations (and hence to illumination related variability). It can be shown~\cite{drumetz2018hyperspectral, uezato2018hyperspectral} that pixelwise endmembers can be defined from the extracted abundances as convex or conical combinations (depending on the constraints used) of the candidates for each class, generalizing the LMM or even the ELMM by considering several signatures per class. A geometric interpretation of unmixing data using bundles with the LMM is shown in Fig.~\ref{bundles_fclsu}~(a).\\
Other purely statistical models can be used, where endmembers are explicitly modeled as drawn from various distributions (Gaussian~\cite{Eches2010} or Mixture of Gaussians~\cite{zhou2017gaussian}, or other variants~\cite{zou2017hyperspectral, halimi2015}), or considered as additively perturbed instances of reference endmembers~\cite{PLMM}. These models, although theoretically able to capture any type of variability, lack the interpretability of physics-based models. An illustration of these models is shown in Fig.~\ref{bundles_fclsu}~(b).
In this paper, we will model intrinsic variability via a simple Bayesian model. Let us assume that we have obtained a model for illumination variability allowing us to define pixelwise endmembers (such as the ELMM). We obtain the mixing model given in Eq.~(\ref{LMM_var}). Since intrinsic variability is really hard to model in a general way, one option is to define it statistically through assumptions on the prior distribution of $\mathbf{S}_{n}$. We model it as the sum of the variability given by the ELMM and a random variable $\mathbf{W}_n \in  \mathbb{R}^{L\times P}$:
\begin{equation}
\mathbf{S}_{n} = \mathbf{S}_{0} \boldsymbol{\psi}_n + \mathbf{W}_n 
\label{model_intrinsic}
\end{equation}
where we can take a simple uniform prior on $\boldsymbol{\psi}_{n}$, while $\mathbf{S}_{0}$ is supposed to have been extracted beforehand. The residual $\mathbf{W}_n$ can be modeled through a Gaussian prior:
\begin{equation}
\mathbf{W}_{n}|\sigma_{\mathbf{S}}^2 \sim \mathcal{N}(\mathbf{0}_{L},\sigma_{\mathbf{S}}^2 \mathbf{I}_{L} )
\end{equation}
In addition, we also assume a spatially and spectrally white Gaussian model error and noise, i.e. 
\begin{equation}
\mathbf{e}_{n}|\sigma_{\mathbf{e}}^2  \sim \mathcal{N}(\mathbf{0},\sigma_{\mathbf{e}}^2  \mathbf{I}_{L} )
\end{equation}
where $\mathbf{I}_{L}\in \mathbb{R}^{L\times L}$ is the identity matrix. If one assumes in addition that the abundances are uniformly distributed in the simplex, then the Maximum A Posteriori (MAP) estimator of the parameters of model~(\ref{LMM_var}) is given, after straightforward computations, by:
\begin{align}
& \underset{\mathbf{A},\mathcal{S},\boldsymbol{\Psi}}{\textrm{argmin}} \ \ \frac{1}{2}  \sum_{n=1}^{N}  \left( ||\mathbf{x}_{n} - \mathbf{S}_{n}\mathbf{a}_{n}||_{2}^{2} + \frac{\sigma_{\mathbf{e}}^2}{2\sigma_{\mathbf{S}}^2} \ ||\mathbf{S}_{n} - \mathbf{S}_{0}\boldsymbol{\psi}_{n}||_{F}^{2} \right)\nonumber  \\
& \textrm{s.t.} \ \ \ \quad \quad \mathbf{a}_{n} \in \Delta_{P} \ \forall n
\label{MAP}
\end{align}
with $\Delta_{P}$ the unit simplex with $P$ vertices.\\
With the physics-based illumination model for extrinsic variability, and the statistical model on intrinsic variability, we obtain a new derivation of the objective function advocated by the papers~\cite{drumetztip, drumetz2015blind}, which first introduced the ELMM.
\section{Analysis of Convex Geometry Concepts for Hyperspectral Unmixing in the presence of endmember variability}
\label{analysis}
The model on endmember variability introduced in the previous section is able to take into account multiple sources of variability, but two major caveats remain:
\begin{itemize}
\item The number of endmembers still has to be estimated in the presence of variability
\item The endmember matrix $\mathbf{S}_{0}$ has to be accurately estimated. Indeed, this parameter is critical since it conditions the whole unmixing chain.
\end{itemize}
In this section, we analyze three different key concepts for hyperspectral image unmixing based on convex geometry when endmember variability is considered, namely ID estimation, endmember extraction and the pure pixel assumption (as well as simplex volume regularization). We describe what endmember variability changes and how it affects a few popular dedicated algorithms designed for in the purely linear case.
%\section{Convex Geometry-based unmixing in the presence of endmember variability}
\vspace{-0.25cm}
\subsection{Intrinsic Dimensionality Estimation}
\label{ID_analysis}
The first step in purely blind unmixing is the determination of the number of endmembers to use. This is typically done by considering that the Intrinsic Dimensionality $d$ (Def.~\ref{ID}) of the data is equal to the number of endmembers. This is true when the LMM holds so long as the endmember matrix $\mathbf{S}$ and the abundance matrix $\mathbf{A}$ have full ranks, since in that case $d = \textrm{dim}(\textrm{span}(\mathbf{SA})) = P$. In the case where illumination-induced variability is present, this result is unchanged since the noiseless data matrix $\mathbf{Y}$ can still be written as $\mathbf{Y} = \mathbf{S}\boldsymbol{\Phi}$ with $\boldsymbol{\Phi}\in \mathbb{R}^{P\times N}$ a coefficient matrix with positive entries. In other words, the span of the conical and convex hulls of $P$ linearly independent points is the same.\\
If we consider in addition intrinsic variability, then the ID of the dataset is likely to change. If all the endmember candidates are stored in a new endmember matrix, the new ID is going to be equal to $d = \textrm{max}(\textrm{rank}(\mathbf{S},\mathbf{A})) \geq P$. Equality can happen, for instance, if all the candidate endmembers corresponding to the same class are the same, or if the abundances of all the instances of each material are zero on the whole support of the image, except for one per endmember. In most cases, the ID of the dataset is then going to be directly related to the total number of possible endmembers. In practice, for real data, where the notions of bundles and candidate endmembers do not really make sense \emph{a priori}, the ID can still provide an upper bound for the number of endmembers, since we can expect pure materials to come in various configurations.\\
With this in mind, all the algorithms estimating ID as the dimension of the signal subspace such as HySIME~~\cite{Bioucas2008} or the Random Matrix Theory (RMT) based algorithm of~\cite{cawse2013} remain useful to give an idea of the number of endmembers to consider. The Virtual Dimensionality (VD) (loosely defined as the ``number of spectrally distinct signatures in the image") concept of~\cite{chang2004} follows the same logic and its value should also increase in the presence of intrinsic variability.
\subsection{Endmember Extraction}
\label{endmember_analysis}
In this section, we analyze how geometrical endmember extraction algorithms based on pure pixel search behave in the presence of variability. Typical convex geometry based endmember extraction algorithms include NFINDR\cite{winter1999n}, the Successive Projection Algorithm (SPA)~\cite{zhang2008successive}, and the Vertex Component Analysis (VCA)~\cite{Nascimento2005}. See~\cite{Plaza2004} for a review. The NFINDR starts from random points in the dataset and iteratively inflates a simplex so as to obtain the simplex with maximum volume that is enclosed in the data. The SPA and VCA (as other algorithms not listed here) are both based on projections of the data on randomly generated vectors, with the assumption that extreme values of these projections are likely to correspond to extreme points in the data scatterplot. In addition, these two algorithms enforce some diversity in the extracted endmembers by projecting the dataset at each iteration to a subspace that is orthogonal to the previously determined endmembers.\\
With a conical model such as the ELMM, endmembers are no longer extreme points in the dataset, but rather lines passing through the origin. However, they can be completely represented by any point (but the origin) on the lines which are the edges of the convex cone. For instance, an endmember is entirely determined by the point of the corresponding line that lies on the unit sphere, since one such point represents a direction in the feature space. It is one representative of an equivalence class for the projective space $\mathbb{RP}^{L}$, defined as the quotient manifold $\left(\mathbb{R}^{L}\backslash \{\mathbf{0} \}\right) / \sim$, where the equivalence relation is $\mathbf{x}\sim \mathbf{y} \Leftrightarrow \exists \psi > 0, \mathbf{x} = \psi \mathbf{y}$. Note that in order to define this space, one has to remove the origin from $\mathbb{R}^{L}$, which suggests that dealing with the origin is going to be problematic with a conical model.\\
With illumination induced variability, the pure pixel assumption amounts to have at least one data point somewhere on the lines defining the endmembers. In~\cite{fu2016robust}, the pure pixel condition is reformulated in a more general NMF identifiability context and referred to as the \emph{separability} condition, stating that at least one coefficient vector (for, say, pixel $n$), per column ($p$) of $\mathbf{S}_0$ in the factorization should be equal to $\psi_{n} \mathbf{z}_{p}$, where $\mathbf{z}_{p}$ is the $p^\textsuperscript{th}$ vector of the canonical basis of $\mathbb{R}^{P}$. If the sum to one constraint is considered in addition, the condition is the same except that $\psi_{n}$ should be equal to one.\\
However, with the classical notion of endmembers as extreme points of the data scatterplot, some extracted spectra would probably (although not in all configurations) include the points on each of those lines associated to the largest scaling factors. The main issue is that with a conical model, the origin itself becomes a salient point of the data, which may lead to the extraction of endmembers with very low amplitude, especially in the presence of shadows or very low brightness materials. Thus, the signal to noise ratio can be extremely low for these points and the resulting endmembers are spurious.\\
Nascimento and Bioucas-Dias, as far back as in 2005, were already conscious about the brightness variations entailed by changing illumination conditions and the possibility to model them as scaling variations~\cite{Nascimento2005-ICA,Nascimento2005}. Thus, they made the VCA algorithm robust to these phenomena by incorporating a \emph{perspective projection}~\cite{Bioucas2013}  step prior to the endmember extraction (this is also sometimes called ``Dark Fixed Point Transform"~\cite{297973}). We define and describe this concept below.
\begin{mydef}
\label{projproj}
Let $\mathcal{C}\subseteq \mathbb{R}^{L}$ be a set and let $\mathbf{u}\in \mathbb{R}^{L}$ be a vector which is not orthogonal to any vector of $\mathcal{C}$. The perspective projection $\mathbf{proj}$ of the set any $\mathbf{x}\in \mathbb{R}^{L}$  onto the hyperplane $\mathbf{x}^\top \mathbf{u} = 1$ is defined as 
\begin{equation}
\mathbf{proj}(\mathbf{x}) = \frac{\mathbf{x}}{\mathbf{x}^{\top}\mathbf{u}}
\end{equation}
\end{mydef}
This projection is not a linear operator, but possesses the property of not being affected by scaling variations. A consequence of this for our unmixing application is the following result, which is already alluded to in\cite{chan2008convex} (Eq.(4)), or~\cite{fu2018nonnegative} (note p.6), but important to bear in mind:
\begin{theorem}
The perspective projection of a polyhedral cone on any compatible hyperplane $\mathbf{u}^\top \mathbf{x} = 1$ is a simplex spanned by the perspective projection of the $P$ generators of the convex cone.
\label{theorem_proj}
\end{theorem}
A proof is provided in the supplementary material to the paper. This theorem means that if the pure pixel hypothesis holds, after applying a perspective projection (VCA uses the mean of the data for $\mathbf{u}$), the data lie in a simplex whose vertices can be identified with any geometrical endmember extraction algorithm. However, in practice, the problem of shadows is not solved, since low brightness pixels are likely to become extreme once projected on the hyperplane. Indeed, in that case, $\mathbf{x}^{\top}\mathbf{u}$ is very small, and hence the norm of the projection is going to be large. This makes these pixels still likely to be selected as endmembers, and as we have already pointed out, they are likely to be spurious because they typically have a low SNR. In real scenarios, problems are also to expect if there are zero or slightly nonnegative values in the data.\\
Another issue is that VCA and other algorithms use random vectors for the projection which are orthogonal to the previously identified endmembers. This makes endmember extraction difficult in cases where we are looking for very correlated endmembers, e.g. different types of vegetation. In summary, geometric endmember extraction algorithms can be used successfully in some cases of illumination induced variability, if a perspective projection step is carried out, and in the absence of very low brightness pixels.\\
If one wants to extract several instances for each endmember to model intrinsic variability through bundles, strategies such as the Automated Endmember Bundles (AEB)~\cite{somers2012} can be used to obtain several instances per material.
\subsection{Pixel Purity and volume regularization}
%\todo[inline]{NO PURE PIXEL CASE - VOLUME REGULARIZATION FOR CONIC MODELS}
The availability of pure pixels is one of the crucial requirements for geometrical endmember extraction algorithms to work in the linear case. However, even in the absence of pure pixels, and in noiseless scenarios, in certain configurations, endmembers can still be perfectly recovered, by finding the simplex enclosing all data points which has minimum volume. The goal of this section is to recall a few results of the literature showing the relevance of resorting to volume regularization when there are no pure pixels, and when in addition endmember variability comes into play.  \\
Note that all the discussion of this section applies only to the noiseless case. The study~\cite{lin2015identifiability} defines the concept of Minimum Volume Enclosing Simplex (MVES) as the largest simplex (in terms of its volume~\cite{gritzmann1995largestj}) which encloses all the data points.\\
In the linear case, when the pure pixel hypothesis is verified, finding the MVES is guaranteed to recover the true endmembers, with little surprise. However, it can be shown that the MVES actually recovers the true endmembers under milder conditions (provided we can find the global optimum of the nonconvex optimization problem -- a highly nontrivial task) involving a quantity called the uniform pixel purity level. In a nutshell, the main result is that the MVES recovers the true simplex if there are enough pixels on the facets of that simplex, even if there are none on the vertices. Another identifiability condition of NMF (although completely different in its formulation) is stated in~\cite{fu2016robust}, defines for a coefficient matrix (the abundance matrix $\mathbf{A}$ or the coefficient matrix $\boldsymbol{\Phi}$, depending on the constraints) the notion of being ``sufficiently scattered". In the same paper, the two conditions are actually shown to be equivalent, even though the latter is less easy to interpret geometrically. Note that these conditions are only sufficient conditions.
%\todo[inline]{ADD one bit about sufficiently scattered condition}
Before moving to the variability case, let us briefly review some algorithms which try to identify the MVES. The optimization problem is nonconvex and involves the determinant of a certain matrix related to the endmembers to compute the volume. This quantity is a hard to work with in optimization. The algorithm of~\cite{5089462}, also referred to as MVES, directly tackles an equivalent formulation of the optimization problem. A different approach is to consider a Nonnegative Matrix Factorization (NMF) technique, where the endmembers are jointly estimated with the abundances, and a volume regularization is added. Examples include the Minimum Volume Constraint NMF (MVC-NMF)~\cite{miao2007endmember}, where the actual volume of the simplex is replaced by a surrogate $\mathrm{det}(\mathbf{S}\mathbf{S}^{\top})$, which is slightly easier to manipulate, because it directly involves the endmember matrix. Other works propose to relax the optimization problem to make it more tractable or robust to noise. The Minimum Volume Simplex Analysis (MVSA)~\cite{li2015minimum} replaces the hard constraint that the simplex must enclose all the data by a soft version to account for outliers. The Iterated Constrained Endmembers (ICE) is another NMF based-algorithm which relaxes the volume computation into a convex surrogate which sums all the pairwise distances between endmembers~\cite{berman2004ice}.\\
Again, a natural question is to wonder how these results can be extended, and the algorithms adapted when endmember variability comes into play. It turns out that the sufficient condition for perfect simplex recovery of the MVES is also valid in the conical case, with some adaptation due to the fact that the sum to one constraint is not enforced. The first necessary step is to define an equivalent to the uniform pixel purity level in the conical case~\cite{fu2016robust}:
\begin{mydef}
The uniform pixel purity level for the extended linear mixing model is a quantity defined as
\begin{equation}
\gamma^{*} = \mathrm{sup} \{r \ | \ \tilde{\mathcal{R}}(r) \subseteq \mathrm{cone}(\mathbf{s}_{1},...,\mathbf{s}_{P})\}
\end{equation}
where 
\begin{align*}
\tilde{\mathcal{R}}(r) & = \{\boldsymbol{\xi}\in \mathrm{cone}(\mathbf{z}_{1},...,\mathbf{z}_{P})\ | \ ||\boldsymbol{\xi}||_2 \leq r\mathds{1}_{P}^{\top}\boldsymbol{\xi} \} \\
& = \{\boldsymbol{\xi}\in \mathbb{R}^{P}\ | \ ||\boldsymbol{\xi}||_2 < r\mathds{1}_{P}^{\top}\boldsymbol{\xi} \} \cap \mathrm{cone}(\mathbf{z}_{1},...,\mathbf{z}_{P})
\numberthis
\end{align*}
\label{gamma_cone}
\end{mydef}
This definition is equivalent to
\begin{equation}
\tilde{\mathcal{R}}(r)  = \left\lbrace \boldsymbol{\xi}\in \mathbb{R}^{P}_{+} \ \middle| \ \frac{\mathds{1}_{P}^{\top}\boldsymbol{\xi}}{||\mathds{1}_{P}||_2||\boldsymbol{\xi}||_{2}} \leq \frac{1}{r\sqrt{P}} \right\rbrace \\ 
\end{equation}
which has a clear geometrical interpretation: $\tilde{\mathcal{R}}(r)$ is simply the set of vectors with nonnegative entries making an angle less than $\textrm{acos}\left(\frac{1}{r\sqrt{P}}\right)$ with the vector $\mathds{1}_{P}$. With this definition, we have the following identifiability result -- a mere reformulation of the criterion proven in~\cite{fu2018identifiability}:
\begin{theorem}
If there are at least three endmembers in the image ($P\geq 3$), and if $\gamma^{*} > \frac{1}{\sqrt{P-1}}$ , then solving the optimization problem
\begin{align}
& \underset{\mathbf{S}\in\mathbb{R}^{L\times P}, \boldsymbol{\Phi} \in \mathbb{R}^{P\times N}}{\mathrm{argmin}} \ \ \mathrm{det}(\mathbf{S}^{\top}\mathbf{S}) \\
& \quad \ \ \ \mathrm{s.t.} \  \quad  \ \  \quad \quad \mathbf{X} = \mathbf{S}\boldsymbol{\Phi} \nonumber \\
 & \quad \ \ \quad \  \  \  \quad  \ \  \quad \quad \boldsymbol{\Phi}\mathds{1}_{N} = \mathds{1}_{P}^{\top} \ \nonumber
\end{align}
guarantees the recovery of the true endmembers (up to scaling factors). In the case $P=2$, this problem recovers the true cone if and only if the pure pixel hypothesis holds.
\end{theorem}
The condition $\mathbf{X} = \mathbf{S}\boldsymbol{\Phi}$ is equivalent to  $\mathbf{x}_{n} \in \mathrm{cone}(\mathbf{s}_{1},...,\mathbf{s}_{P}), \ \forall n = 1,...,N$, but this time the coefficient matrix has to appear explicitly, because there is a sum to one constraint on its \emph{rows} (not columns, which would be the usual sum to one constraint on an abundance matrix). This constraint is only a technical constraint which ensures that minimizing the cost function leads to identify the edges of the cone. However, it breaks the geometrical interpretation of the model since the objective function cannot be interpreted anymore as the volume of a simplex such that the conical hull of its vertices enclose the data. However, without more constraints on the columns of $\mathbf{S}$ (e.g. unit norm, which would make it impossible to satisfy the sum to one constraint on the rows of $\boldsymbol{\Psi}$) this quantity cannot be easily interpreted as an extension of the volume of a simplex to the associated polyhedral cone (such a measure should have some sort of scale invariance with respect to the columns of $\mathbf{S}$).
%However, one can get a geometrical intuition of why the criterion $\mathrm{det}(\mathbf{S}\mathbf{S})^{\top}$ is a measure of the size of the cone spanned by the columns of $\mathbf{S}$ since $\mathrm{det}(\mathbf{S}\boldsymbol{\psi} (\mathbf{S}\boldsymbol{\psi})^{\top}) = \prod_{i=1}^{P} \psi_{i}$ and the criterion is higher if one scales the 
Once again, this criterion is only a sufficient condition, but empirical evidence suggests that it could also be a necessary condition for identifiability~\cite{fu2018nonnegative}. However, those results make us confident that volume regularization techniques could still help identify the lines of the endmembers in the case of illumination variability in practical scenarios, even in the absence of pure pixels.\\
When, in addition, we consider intrinsic variability, the analysis does not apply anymore. However, even if we consider intrinsic variability without illumination-induced variability, we are not so much interested in the extreme points of the data, as in the centroids of the convex hulls of each bundle, which allow us to define good reference endmembers. With a conical model, the rationale is the same, we do not want the edges of the convex cone anymore, but rather the center of the conical hulls of the candidate endmembers for each material. Hence the volume regularization still makes sense, but it might be preferable to resort to a minimum volume based algorithm with a soft constraint on the inclusion of the whole data in the cone, such as MVSA or ICE, simply adding the unit norm constraint on the reference endmembers.
\section{Proposed Unmixing Framework}
\label{chain}
%\todo[inline]{EXPLAIN ICASSP ALGORITHM - START FROM THE MODEL OF EQ 4 - AND DESCRIBE THE ALGORITHM WITH A BAYESIAN POINT OF VIEW, GAUSSIAN PRIOR ON THE LOCAL ENDMEMBERS, CENTERED ON THE LINES OF THE ELMM (both illumination and intrinsic var are modeled), NECESSITY OF OBLIQUE MANIFOLD CONSTRAINT, K MEANS INITIALIZATION (INTERPRETATION AS A MIXTURE OF VMF DISTRIBUTIONS) AND VOLUME REGULARIZATION ON THE ENDMEMBERS (EQUIVALENCE WITH WATSON PRIOR ON THE REFERENCE ENDMEMBERS)}
In this section, with all the analyses of the previous sections in mind, we outline the core contribution of this paper, a full unmixing chain which applies all convex geometry concepts of linear unmixing to the case where both types of variability can be found in the data. The first step is to estimate an upper bound of the number of the dataset through ID estimation, using one's favorite algorithm. The second step is endmember extraction to define reference endmembers, and the last one is abundance, refined reference endmember and variability estimation. This last step is an NMF step which is similar to the approach of~\cite{drumetztip}, except that volume regularization is added in a way that is compatible with conical data, and which is robust to intrinsic variability. We describe the reference endmember extraction and the NMF step below.
\subsection{Reference endmember extraction}
We aim at extracting reference endmembers around which variability can be extrapolated on each pixel of the data. Even though both intrinsic variability and illumination-induced variability can be expected in real scenarios, in the absence of shadows and low brightness pixels, the perspective projection step of the VCA makes it a good candidate algorithm to use. However, if shadows or outliers are an issue, another way of extracting reference endmembers should be used. In such cases, we propose to cluster the data using a simple k-means algorithm with $P$ clusters, and the spectral angle as the similarity measure. We use the centroids as initial reference endmembers, which will then be refined in the next step.\\
The reason to resort to k-means with the cosine distance, instead of a geometrical extraction endmbmer is that it allows us to define the endmembers taking into account the statistics of the data, as well as their geometry. We expect to be able to capture correlated endmembers more easily than with the VCA, which, to identify a new endmember, iteratively projects the data on orthogonal vectors to the subspace spanned by the already identified endmembers (see~\cite{Nascimento2005} for details). Also, low brightness pixels are not given as much importance as with the VCA. Another theoretical motivation to use this strategy is that, as the regular k-means with the Euclidean distance can be interpreted as a hard assignment version of a Gaussian Mixture Model, the k-means algorithm can be interpreted as a hard assignment version of a mixture of Von Mises-Fisher distributions~\cite{banerjee2005clustering}. This distribution can be thought of as the equivalent of an isotropic Gaussian distribution on the sphere, i.e. for directional data, such as endmembers in a conical model~\cite{mardia2009directional}. The main drawback of using k-means is that it performs a hard assignment, and hence does not account for mixed pixels, and as a result, the centroids of the clusters can be too far in the conical hull of the data because very mixed pixels have to be assigned to one class only. This is why it is necessary to be able to adjust the endmembers during the abundance estimation step. The pure pixel assumption helps, but since the refrerence endmembers will be adjusted using a volume related criterion, it is not strictly mandatory in theory.
\subsection{NMF estimation of the abundances, reference endmembers and variability}
\label{absest}
At this step, we have initial estimates for the reference endmembers, and we need to estimate the abundances, as well an endmember matrix for each pixel, accounting for both types of variability. We first define the cost function that we are going to minimize, by slightly extending the Bayesian variability model of Eq.~(\ref{model_intrinsic}). Since our goal is to reestimate $\mathbf{S}_{0}$ adaptively, we include the following  hyperprior on the reference endmembers: 
\begin{equation}
\mathbf{S}_{0}|\delta \propto \exp (-\delta\mathrm{tr} (\mathbf{S}_{0}\mathbf{V}\mathbf{S}_{0}^{\top})), \ \textrm{with} \ \mathbf{S}_{0} \in \mathcal{OB}(L,P).
\end{equation}
where $\textrm{tr}$ denotes the trace of a matrix, and $\mathbf{V} = P \mathbf{I}_{P} - \mathds{1}_{P}\mathds{1}_{P}^{\top}$, such that
\begin{equation}
\frac{1}{2}\ \textrm{tr}  (\mathbf{S}_{0}\mathbf{V}\mathbf{S}_{0}^{\top}) = \sum_{i=1}^{P-1}\sum_{j=i+1}^{P} ||\mathbf{s}_{0i} - \mathbf{s}_{0j}||^{2}_{2},
\end{equation} i.e. the sum of pairwise Euclidean distances between reference endmembers. A derivation of this equality can be found in~\cite{berman2004ice}. $\mathcal{OB}(L,P)$ denotes the \emph{oblique manifold}, i.e. the manifold of matrices with unit norm columns. With this additional hyperprior, the MAP estimator of the model~\eqref{model_intrinsic} becomes
\begin{align}
& \underset{\mathbf{A},\mathcal{S},\boldsymbol{\Psi},\mathbf{S}_{0}}{\textrm{argmin}} \ \ \frac{1}{2}  \sum_{n=1}^{N}  \left( ||\mathbf{x}_{n} - \mathbf{S}_{n}\mathbf{a}_{n}||_{2}^{2} + \lambda_{S} \ ||\mathbf{S}_{n} - \mathbf{S}_{0}\boldsymbol{\psi}_{n}||_{F}^{2} \right)\nonumber  \\
 & \quad \quad \quad \quad + \frac{\lambda_{S_{0}}}{2} \  \textrm{tr} (\mathbf{S_{0}}\mathbf{V}\mathbf{S_{0}^{\top}})\nonumber \\
& \textrm{s.t.} \ \ \ \quad \quad \mathbf{a}_{n} \in \Delta_{P} \ \forall n \nonumber \\
 &  \quad \quad \quad \quad \mathbf{S}_{0} \in \mathcal{OB}(L,P)
\label{objective}
\end{align}
where we have replaced the estimation of the standard deviations and of $\delta$ with regularization parameters to tune. \\
The new additional term allows to control the openness of the cone (depending on the value of $\lambda_{S_{0}}$), with the convex approximation of the volume used in the ICE algorithm~\cite{berman2004ice}. This term can really be thought of as an approximation for the volume of the cone, and we further constrain the endmembers to be normalized, treating them as \emph{directional data}~\cite{mardia2009directional}. Without this, we compare endmembers which could have different magnitudes, which is not meaningful. The openness of the cone can then be tuned so that the reference endmembers are situated within the data (in the presence of pure pixels and with intrinsic variability), or outside of the dataset (if there are no pure pixels). The term $\lambda_{S} \ ||(\mathbf{S}_{n} - \mathbf{S}_{0}\boldsymbol{\psi}_{n})||_{F}^{2}$, forces each local endmember to be close (but not equal) to scaled versions the (unit norm) representatives of the reference directions. The scaling factors capture illumination induced variability, while $\mathbf{S}_{n}$ can further account for intrinsic variability effects, by allowing the local endmembers to drift away from the lines, depending on the value of $\lambda_S$. The fact that the reference endmembers are normalized also has the advantage of easily allowing to compare the magnitude of the scaling factors (and thus the impact of illumination induced variability) across different materials and images. Spatial regularizations could also be added if need be, as done in~\cite{drumetztip}. The novelty of the proposed criterion relies on the reference endmember term, together with the oblique manifold constraint to bridge the hitherto separate issues of volume regularization and endmember variability.\\
\subsubsection{Optimization}
Here, we propose an algorithm to obtain a stationary point of the cost function~\eqref{objective}. This objective function is challenging for several reasons: it is nonconvex over all variables simultaneously, which usually calls for block coordinate descent methods to try and get a local minimum. In this case, this approach is made even more complex because the problem is not convex w.r.t. $\mathbf{S_{0}}$ either, because of the nonconvex unit norm constraints. However, we will see we can still obtain a local minimum for this variable by taking advantage of the Riemannian manifold structure of the constraint set. Before detailing the different steps of the optimization, we will briefly describe how we initialize the algorithm. As mentioned above, if VCA is not used for the endmember extraction, we first run the k-means clustering algorithm (with the cosine similarity) to obtain centroids, which we normalize to initialize $\mathbf{S}_{0}$. We initialize $\mathbf{S}_{n}$ by assigning the appropriate column of this matrix to the current pixel $\mathbf{x}_{n}$, depending on its clustering label. The other columns are initialized using the remaining centroids. The abundance and scaling factor matrices are initialized using the SCLSU algorithm with the centroids as references, which is very fast. This way, we hope to obtain a good local minimum in spite of the complexity of the problem.\\
The optimization w.r.t. $\mathbf{A}$ is relatively easy, since the objective function is smooth, convex and the constraint set (unit simplex) is easy to project onto~\cite{condat2014fast}. The global minimum of this subproblem can be then obtained pixel-by-pixel using (for instance) a projected gradient descent. The optimizations w.r.t to $\mathbf{S}_{n}$ and $\boldsymbol{\psi}_{n}$ are easy and enjoy closed form solutions (see~\cite{drumetz2015blind} for details). Optimizing over $\mathbf{S}_{0}$ is harder because of the unit sphere constraints, despite the smoothness and convexity of the objective. Using the fact that the constraint set has a Riemannian manifold structure for which a retraction mapping can be easily found, we perform a conjugate gradient descent on the oblique manifold~\cite{absil2009optimization} (we use the Manopt toolbox for MATLAB~\cite{manopt}-- also available in Python).\\
%Due to the nonconvex nature of the oblique manifold constraint, we cannot prove global convergence, even though we observe good performance in practice. If this constraint is dropped, the objective is globally smooth and multiconvex. Unfotunately even in that case, we are not able to guarantee that the initial strong convexity of the cost function w.r.t. to each block remains valid along all the iterations. This property would ensure global convergence~\cite{xu2013block}. Convergence could be achieved by resorting to inexact Block Coordinate Descent schemes instead of exact minimizations, as done in~\cite{xu2013block}.\\
Due to the nonconvex nature of the oblique manifold constraint, we cannot prove global convergence, even though we observe good performance in practice. However, if this constraint is dropped, we can find a closed form update for $\mathbf{S}_{0}$, and we can prove convergence of the algorithm to a stationary point. We use the result of~\cite{grippo2000convergence} (Proposition 5) on block coordinate descent (BCD) techniques. This results states that if the objective function is continuously differentiable over the global optimization variable, is strictly quasiconvex w.r.t. $m-2$ of the $m$ considered blocks, and if the constraint set is convex, then BCD converges to a stationary point of the objective. In our case, our optimization algorithm can be seen as a 3-block BCD, using blocks $\mathcal{S}$, $\{ \mathbf{A},\boldsymbol{\Psi} \}$ and $\mathbf{S}_{0}$. The objective function is obviously continuously differentiable, and is in addition strictly convex w.r.t. $\mathcal{S}$ (so long as $\lambda_{S} > 0$). Indeed, for each $n$, if we consider the variable to be $\textrm{vec}(\mathbf{S}_{n}) \in \mathbb{R}^{LP}$ (stacking the columns), the Hessian is equal to $(\mathbf{a}_{n} \mathbf{a}_{n}^{\top}) \otimes \mathbf{I}_{L} + \lambda_{S} \mathbf{I}_{L}\otimes \mathbf{I}_{P} \in \mathbb{R}^{LP\times LP}$, where $\otimes$ is the Kronecker product. This is a symmetric positive definite matrix for $\lambda_S > 0$, and then the objective is \emph{a fortiori} quasiconvex w.r.t. $\mathcal{S}$. Since we dropped the nonconvex normalization constraint, the result applies and the algorithm converges in that particular case.\\
%Finally, note that we could also obtain a global convergence of the whole process, should we be willing to normalize the reference endmembers using an $\mathcal{L}_{1}$ norm, and further constraining the endmembers to have nonnegative entries, instead of just using the Euclidean norm. In this case, the constraint set is a simplex and is convex. However, using this normalization makes us lose the intepretation of endmembers as directional data, and we do not use it here.\\
% detail a bit the gradient descent on the manifold if enough space
We stop the algorithm whenever the relative variations between consecutive iterates of all blocks of variables go below $\epsilon = 10^{-3}$ (in norms). We note that the convergence is going to be slower than the original ELMM with fixed reference endmembers, because the latter are now iteratively updated and impact the whole geometry of the unmixing.
\vspace{-0.15cm}
\section{Experimental Results}
\label{results}
\vspace{-0.1cm}
In this section, we present the unmixing results obtained on a synthetic dataset whose materials incorporate realistic variability features, as well as a challenging real dataset with very correlated endembers and the presence of a significant proportion of shadowed areas.
\subsection{Synthetic data}
\subsubsection{Data Generation}
To generate a synthetic dataset, we first use the ground truth of the well-known Pavia University dataset to provide us with labeled spectra (203 bands in the visible and near infrared domains) belonging to several classes of interest, incorporating their spectral variability. We consider three classes present in the image: vegetation, concrete and metallic roofs. Theses classes incorporate both illumination induced variability (roofs and trees locally have different orientations w.r.t. the sun) and more intrinsic variability sources (especially in concrete and vegetation). In each pixel, we choose the local endmembers to be a random sample within each of these classes (after a normalization to project them on the unit sphere).\\
Scaling factors have been simulated by drawing them from a mixture of 4 Gaussian distributions (fitted from the results of SCLSU on a subimage of the Pavia dataset), which reflects the fact that in real scenarios scaling factors often come from multimodal distributions (for example roofs with two different orientations, or areas with shadows).\\
The abundances have been designed to be relatively sparse, using a Dirichlet distribution such that the probability density is concentrated around the edges and vertices of the unit simplex (while still allowing a proportion of heavily mixed pixels), so in this case the pure pixel assumption holds.\\
The data was then generated using Eq.~\eqref{LMM_var_mat}, adding Gaussian white (both spatially and spectrally) noise such that the signal to noise ratio is 30dB. The generated image then benefits from realistic statistical properties.
\subsubsection{Results}
\begin{figure}
\begin{center}
{\includegraphics[scale=0.25]{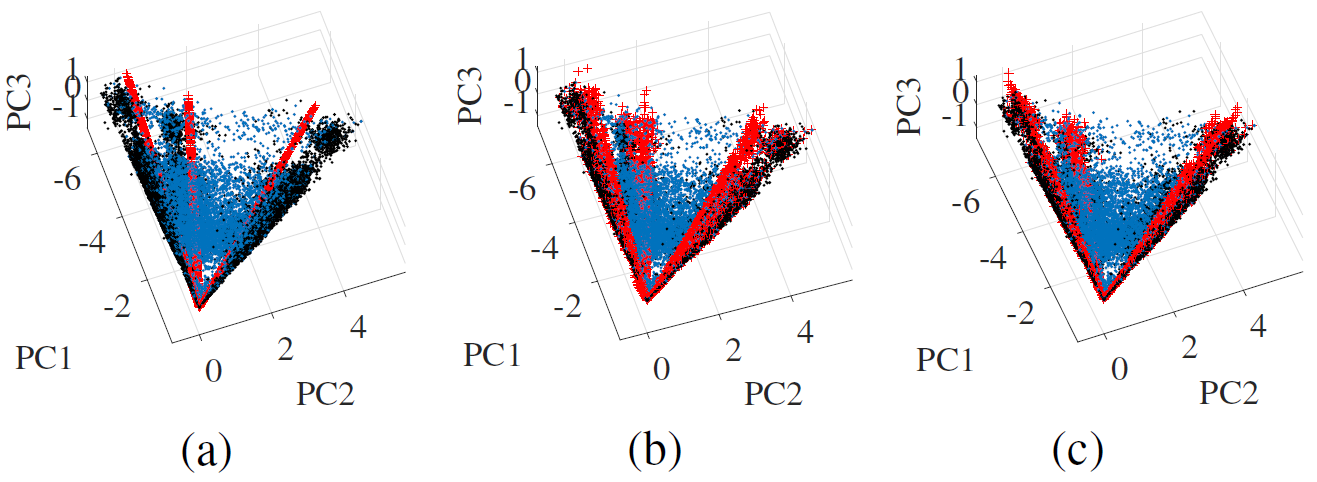}}
\end{center}
\vspace{-0.5cm}
\caption{Scatterplots of the data (blue), the true endmembers (black) and the extracted ones (red) for (a) SCLSU  (b) ELMM (c) RELMM.}
\label{synthetic_scatterplots}
\vspace{-0.25cm}
\end{figure}

First, we ran the HySIME~\cite{Bioucas2008} and the RMT~\cite{cawse2013} algorithm on this dataset, which gave an ID value of 14, and 29, respectively, even though only 3 endmembers are considered. The reason for this is that all the endmember variants introduced in the data generation lead to a substantial increase in the ID value, as explained in Sec.~\ref{ID_analysis}, and this experimentally confirms that the ID can be considered more as an upper bound of the number of endmembers, rather than an absolute truth.\\
We run the the SCLSU algorithm with VCA-derived endmembers, so as to get baseline unmixing results with variability. We show below that this approach fails in this configuration. Then we focus on testing two algorithms with k-means derived references: SCLSU and the ELMM algorithm, as presented in~\cite{drumetz2015blind}. Also, we denote by ELMM-SSD (Sum of Squared Distances) the ELMM augmented with the convex volume regularization of the ICE algorithm, but without the oblique manifold constraint. Finally, we compare all those methods to one proposed in Sec.~\ref{absest}, denoted as RELMM (for Robust ELMM). Note that we do not compare the results to the classical Fully Constrained Least Squares Unmixing~\cite{Heinz2001}, because this algorithm assumes a simplex-based model and has been shown to fail in many endmember variability scenarios. For each algorithm, we empirically tune the regularization parameters to obtain the best possible performance (the chosen values are reported in Table~\ref{table}). Quantitative results are presented using two metrics: the abundance Root Mean Squared Error (aRMSE) between the true abundances and the recovered ones: $\frac{1}{N\sqrt{P}}\sum_{n=1}^{N} ||\hat{\mathbf{a}}_{n}-\mathbf{a}_{n}||_{2}$, and the mean (over all pixels and materials) Spectral Angle Mapper (SAM) (Eq.~\eqref{SAM}) between the true endmembers in each pixel and the recovered ones. These quantities are gathered in Table~\ref{table}.
\begin{table}
\begin{center}
\scalebox{0.9}{
\begin{tabular}{|c|c|c|c|c|c|}                          
\hline 
                        & $\lambda_{S}$ & $\lambda_{S_{0}}$  & aRMSE            &  SAM (degrees) & Time (s)     \\
\hline 
VCA+SCLSU &  $\times$ & $\times$ &  0.2075 & 54.4 &  3\\ 
\hline
SCLSU  & $\times$ & $\times$& 0.0654 & 6.32 & 2 \\
 \hline
ELMM  & 0.01 & $\times$ & 0.0642 & 5.62 & 18\\ 
\hline
ELMM+SSD & 0.1 & 0.25 & 0.1718  & 10.41  & 88 \\       
\hline
RELMM  & 0.1 & 0.5 & \textbf{0.0560}&  \textbf{3.48}  & 428 \\
\hline
\end{tabular}}
\end{center}
\caption{Quantitative results on the synthetic data. Except for VCA+SLCSU, all algorithms use k-means to obtain the initial reference endmember matrix. Regularization parameters values are reported when applicable.}
\label{table}
\vspace{-1cm}
\end{table}
VCA+SCLSU obtains very poor results both in abundance estimation and variability retrieval, because two of the extracted signatures are associated with pixels with small scaling factors, and have a very low magnitude. The reason for this behavior is explained in Sec~\ref{endmember_analysis}. Using k-means instead along with SCLSU leads to better results, but far from optimal because the variability is only explained by scaling factors, and hence intrinsic variability is not accounted for. The ELMM does even better because it addresses it using an additional Gaussian prior on the local endmembers, allowing them to drift away from the lines defining the reference endmembers. ELMM+SSD fails because the regularization term involves the comparison of references with possibly different scales, whereas introducing the constraint leads to the best results.\\
Fig.~\ref{synthetic_scatterplots} shows qualitative results using 3D-scatterplots of the data (using the first principal components) along with the recovered and true endmembers for the best algorithms (for the other algorithms, the lines are much too far away from the true cone to be relevant). Similar conclusions can be drawn from this figure, showing that RELMM finds the best endmembers in each pixel. Geometrically, the local endmembers defined using the volume regularization allow us to position the references so they lie more or less in the center of cones spanned by the true endmembers corresponding to each class. The k-means endmembers are initially slightly too far within the data and need to be updated to improve the unmixing performance.
\vspace{-0.1cm}
\subsection{Real data}
\begin{figure}
\begin{minipage}{0.32\linewidth}
  \centering
{\includegraphics[scale=0.185]{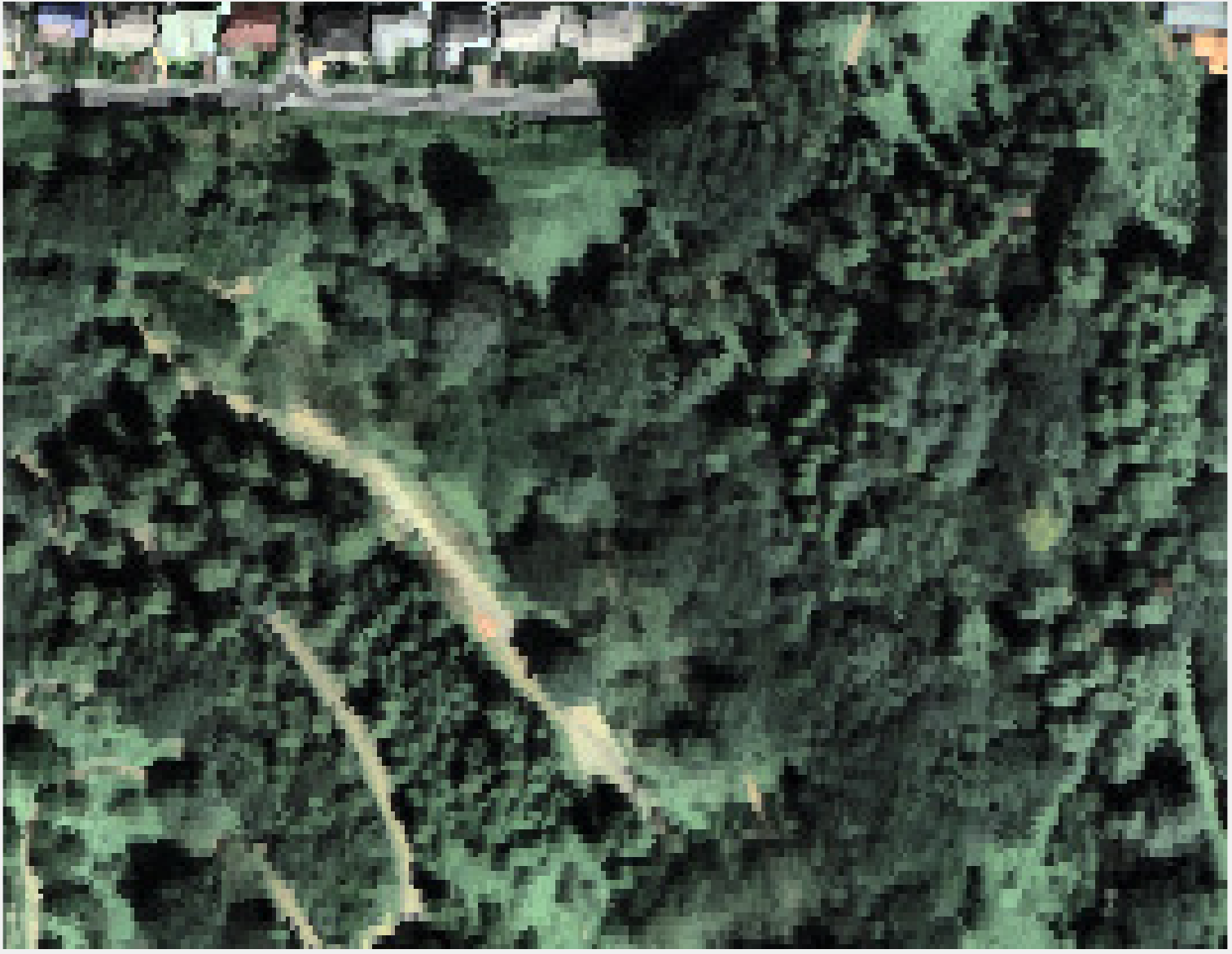}}
  \centerline{(a)}\medskip
  \end{minipage}
  \hspace{0.05cm}
\begin{minipage}{0.32\linewidth}
  \centering
\includegraphics[scale=0.4]{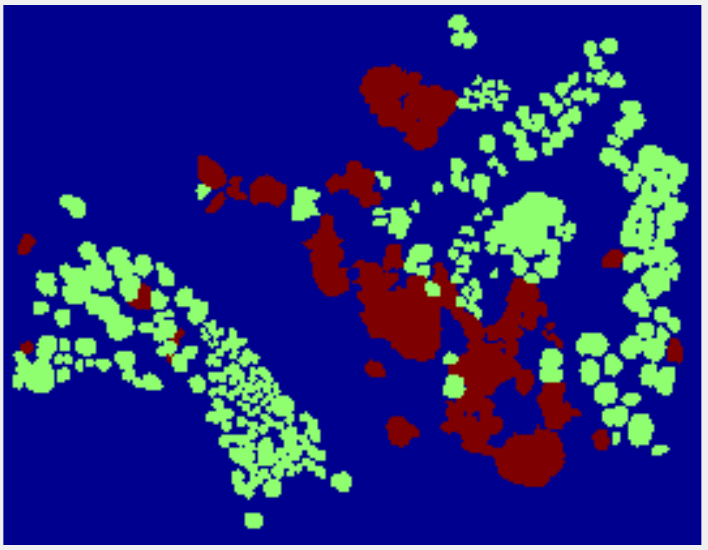}
  \centerline{(b)}\medskip
\end{minipage}
\hspace{0.05cm}
\begin{minipage}{0.3\linewidth}
  \centering
\includegraphics[scale=0.22]{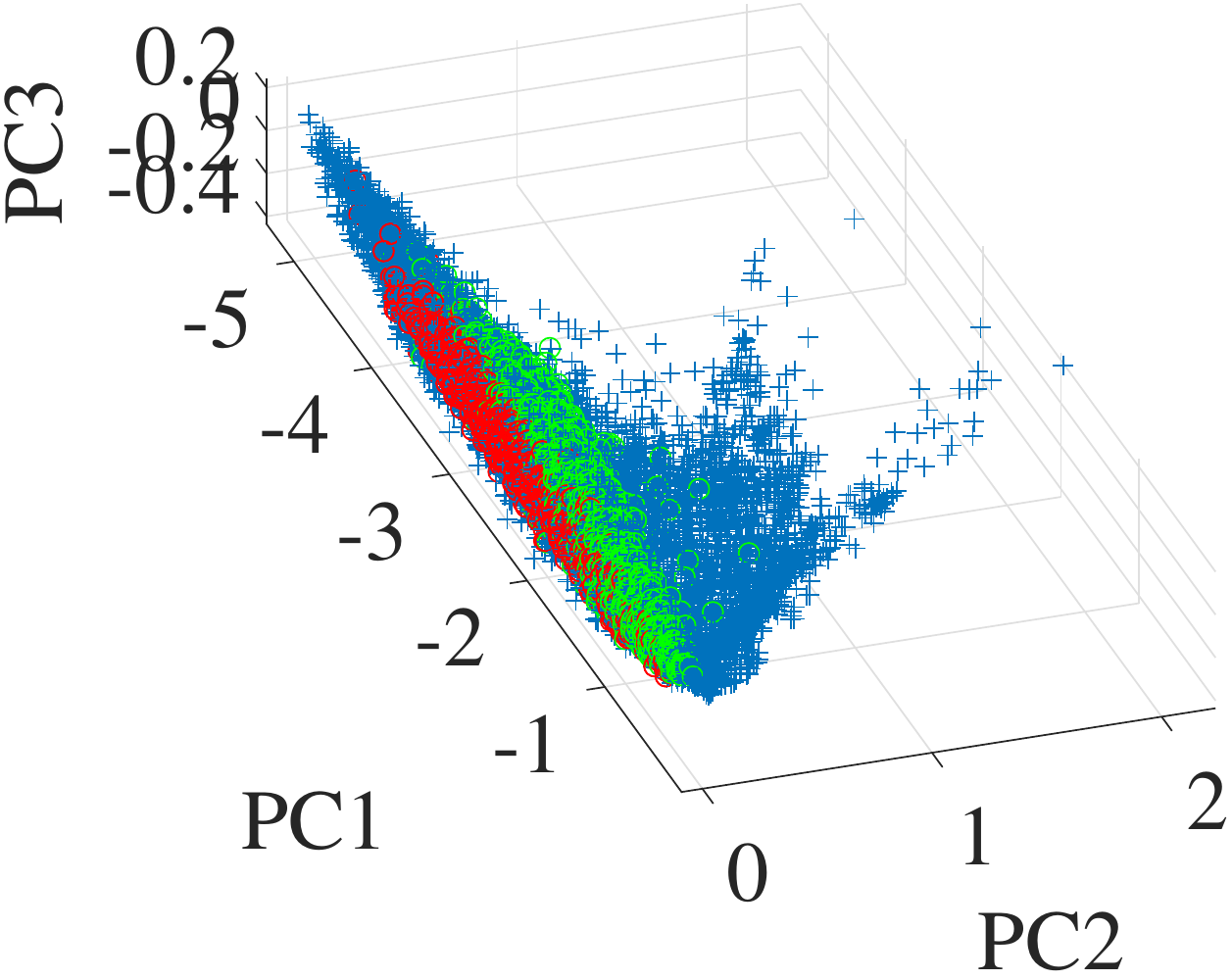}
  \centerline{(c)}\medskip
\end{minipage}
\vspace{-0.5cm}
\caption{(a) RGB representation of the data. (b) Ground truth for conifer (green) and broadleaf (red) trees. (c) Scatterplot of the data, as well as the ground truth (same color code).}
\label{tama_forest}
\vspace{-0.2cm}
\end{figure}

\begin{figure*}
\begin{center}
\includegraphics[scale=0.5]{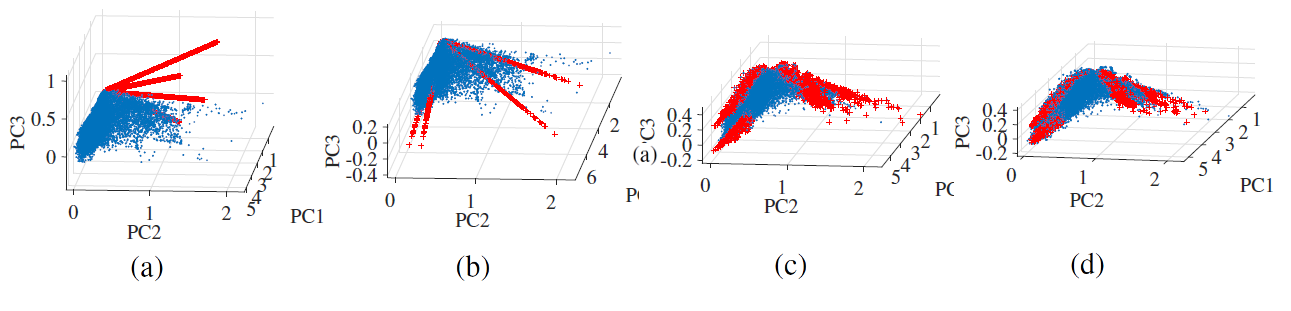}
\vspace{-0.5cm}
\caption{Scatterplots of the data (blue) and the extracted endmembers (red) for (a) VCA+SCLSU (b) SCLSU  (c) ELMM (d) RELMM.}
\label{real_scatterplots}
\end{center}
\end{figure*}

\begin{figure*}
  \centering
\begin{minipage}{0.59\linewidth}
{\includegraphics[scale=0.7]{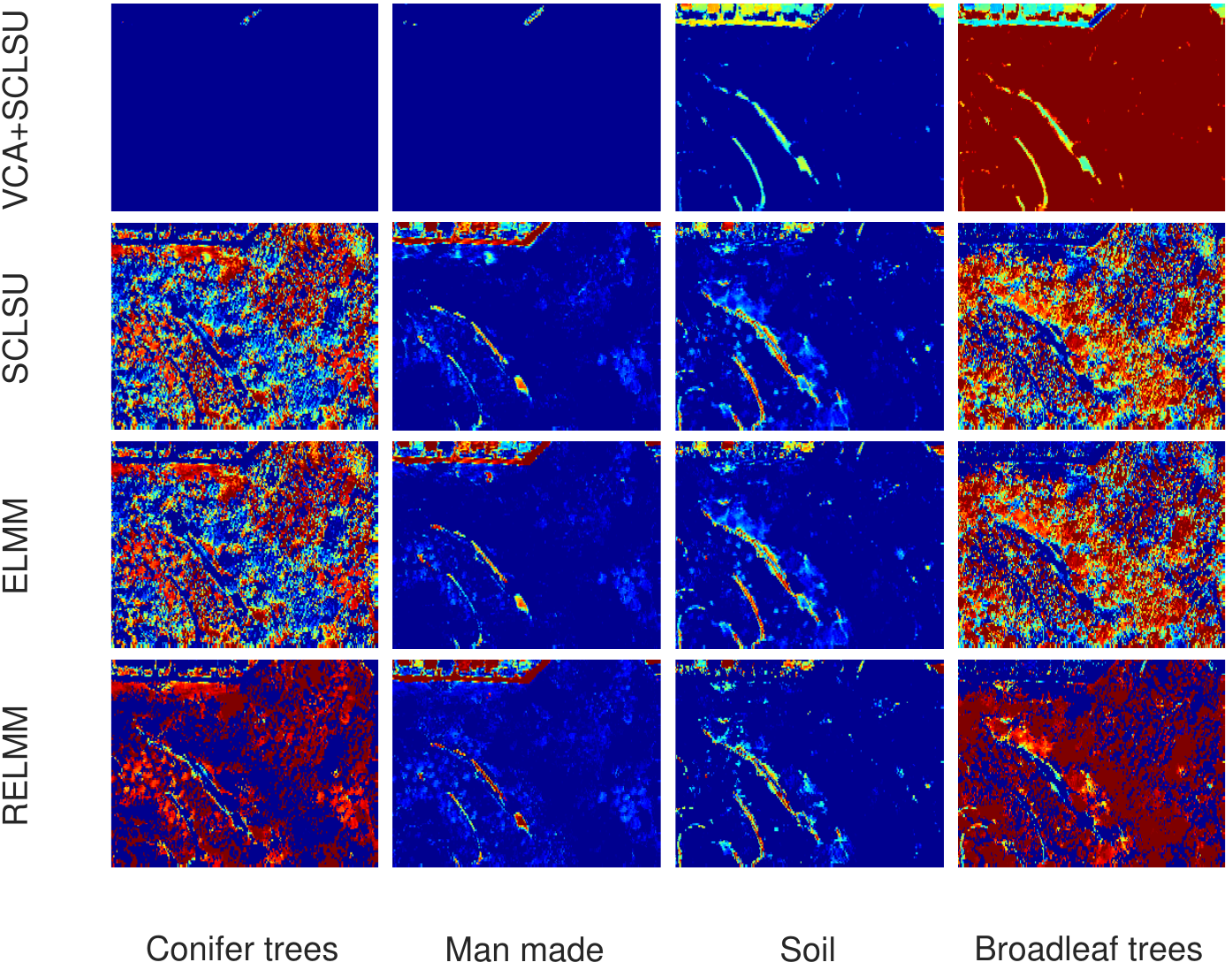} \includegraphics[scale=0.125]{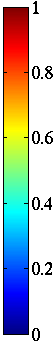}}
  \centerline{(a)}\medskip
\end{minipage}
  \centering
\begin{minipage}{0.39\linewidth}
{\includegraphics[scale=0.6]{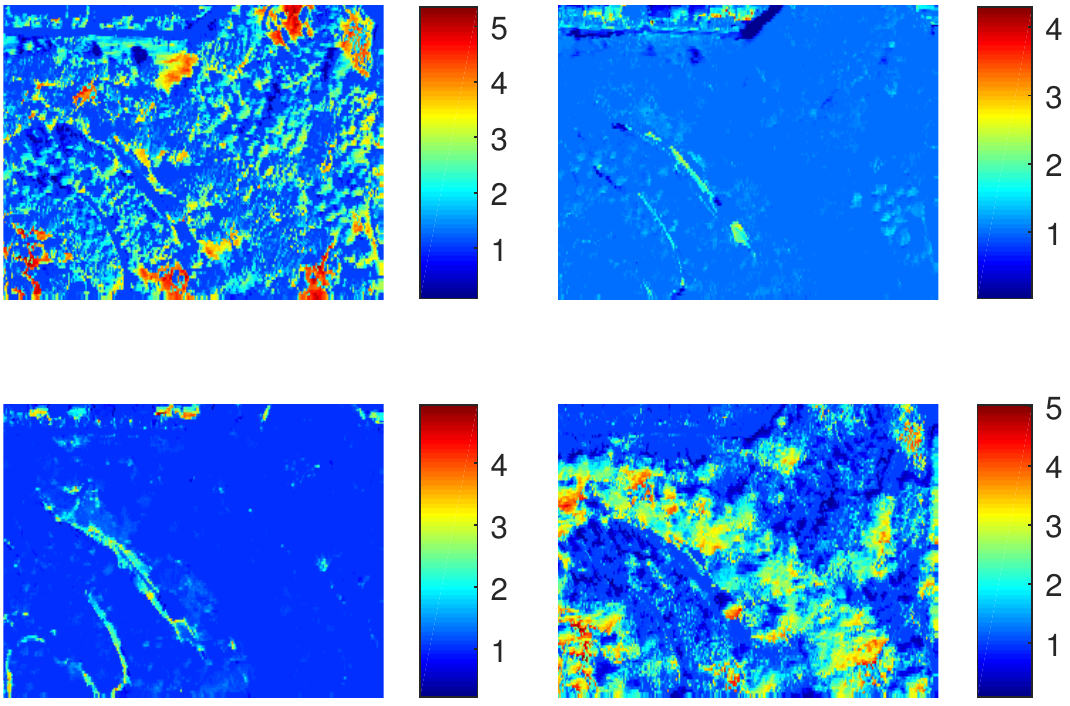}}
  \centerline{(b)}\medskip
\end{minipage}
\vspace{-0.25cm}
\caption{(a) Abundances maps obtained by the tested algorithms. (b) Scaling factors obtained by RELMM. Note how large values in the conifer trees actually account for grass in the image.}
\label{real_res}
\end{figure*}

The real dataset we use in this study was acquired in 2009 by Japan Space Systems over the Tama Forest Science Garden in the western region of Tokyo, Japan, with the CASI-3 sensor (72 spectral bands in the visible and near-infrared domains)~\cite{matsuki2015hyperspectral}. The spatial resolution is 1m. The image we use is a $207\times 268\times 72$ subset of the whole scene. An RGB representation is shown in Fig.~\ref{tama_forest} (a). This dataset has been used for supervised classification of tree species, using a ground truth and LiDAR (Light Detection And Ranging) data as an additional classification feature, since the different tree species are spectrally very close to one another. The image also comprises many shadowed areas because of the tree crowns, which were an important hurdle in previous studies~\cite{matsuki2015hyperspectral}. Furthermore, other non vegetation endmembers are present, such as man made roofs, roads, and soil. We show here that using k-means instead of the VCA allows to distinguish between conifer and broadleaf trees in a completely unsupervised way. Some labeled conifer and broadleaf trees are shown in Fig.~\ref{tama_forest} (b). We show the scatterplot of the data and labeled pure pixels in Fig.~\ref{tama_forest}(c).
%\begin{figure}
%\begin{minipage}{0.32\linewidth}
%  \centering
%{\includegraphics[scale=0.185]{imadjust_rgb.pdf}}
%  \centerline{(a)}\medskip
%  \end{minipage}
%  \hspace{0.05cm}
%\begin{minipage}{0.32\linewidth}
%  \centering
%\includegraphics[scale=0.4]{gt_tree_types.pdf}
%  \centerline{(b)}\medskip
%\end{minipage}
%\hspace{0.05cm}
%\begin{minipage}{0.3\linewidth}
%  \centering
%\includegraphics[scale=0.22]{pca_classes_scatter.pdf}
%  \centerline{(c)}\medskip
%\end{minipage}
%\caption{(a) RGB representation of the data. (b) Ground truth for conifer (green) and broadleaf (red) trees. (c) Scatterplot of the data, as well as the ground truth (same color code).}
%\label{tama_forest}
%\end{figure}
The HySIME algorithm estimated the ID of the data to be 14, the RMT algorithm estimated it the ID to be 21. Because of the important variability present in the image, and since we may want to group several macroscopic constituents of the man-made material into one endmember for easier unmixing an interpretation (roads, several types of roofs...), we unmix the data using $P = 4$ materials using the above mentioned algorithms. We show in Fig.~\ref{real_scatterplots} the scatterplots of the data and recovered endmembers for VCA+SCLSU, SCLSU, ELMM, and RELMM. The abundance maps are shown in Fig.~\ref{real_res}~(a). For the ELMM, we set $\lambda_{S} = 0.01$, and for the RELMM, we set $\lambda_{S} = 0.5$ and $\lambda_{S_{0}} = 100$. As in the synthetic data case, the endmembers recovered by VCA are spurious because of shadow patches of the image, and the corresponding abundances are meaningless. Most of the data is projected on the closest line in the identified cone, which represents vegetation. Using k-means instead allows to distinguish between conifer trees and broadleaf trees. Grass and shadows are also detected by large and low values of the scaling factors, respectively (see Fig~\ref{real_res}~(b)). The abundances of SCLSU and the ELMM are rather similar, slightly sparser for the ELMM, because it is able to better capture variability effects than SCLU (as seen in Fig.~\ref{real_scatterplots}~(c). The RELMM, thanks to being able to adjust the references, is able to obtain sparser abundance maps which closely match the ground truth of Fig.~\ref{tama_forest} (b). We see that the identified endmembers enclose the data very well and are the closest to the ground truth pixels of Fig.~\ref{tama_forest}~(c). The values of the regularization parameters used and the running times are provided in Table~\ref{table1}.
\begin{table}
\begin{center}
\scalebox{0.85}{
\begin{tabular}{|c|c|c|c|}                          
\hline 
                        & $\lambda_{S}$ & $\lambda_{S_{0}}$  & Time (s)     \\
\hline 
VCA+SCLSU &  $\times$ & $\times$  &  9\\ 
\hline
SCLSU  & $\times$ & $\times$ & 11 \\
 \hline
ELMM  & 0.01 & $\times$ & 243\\      
\hline
RELMM  & 0.5 & 100 & 757 \\
\hline
\end{tabular}
}
\end{center}
\caption{Results on the real dataset. Except VCA+SLCSU, all algorithms use k-means to obtain the initial reference endmember matrix.}
\vspace{-0.7cm}
\label{table1}
\end{table}
%\begin{figure*}
%  \centering
%{\includegraphics[scale=0.9]{abs_real_first.pdf}}
%\includegraphics[scale=0.125]{abs_legend.png}
%\caption{Abundances maps obtained by the tested algorithms.}
%\label{real_abs}
%\end{figure*}
%\begin{figure}
%  \centering
%{\includegraphics[scale=0.8]{psis_RELMM.pdf}}
%\caption{Scaling factors obtained by RELMM. Note how large values in the conifer trees actually account for grass in the image.}
%\label{psis}
%\end{figure}
\vspace{-0.4cm}
\section{Conclusion}
\label{conclusion}
Blind unmixing is a problem of prime importance for the analysis of hyperspectral images. The classical linear unmixing chain relies heavily on convex geometry concepts to extract endmembers and abundances, but many recent works suggest that endmember variability should be taken into account for accurate unmixing. Hence, the geometrical assumptions of the linear mixing model have to be adapted. We analyzed the main steps of the typical unmixing chain, namely the estimation of the number of endmembers, and their extraction in the presence of pure pixels or not, as well as abundance estimation. We modeled variability in two complementary ways: illumination induced variability was modeled through the extended linear mixing model, allowing a local scaling of the endmembers, and intrinsic variability effects were modeled using a statistical approach. We found that the estimation of the Intrinsic Dimensionality of the data can theoretically still provide an upper bound of the number of endmembers to use, that the VCA algorithm is robust to illumination induced variability provided there are no very low britghness pixels in the image. Otherwise, we advocate the use of k-means clustering with the cosine similarity to obtain initial estimates of the endmembers. Both algorithms can also be used to extract endmember bundles. By gathering various results of the literature, we have extended the notion of pure pixels and minimum volume identifiability results to conical mixing models. We have also stressed that in the presence of pure pixels, these approaches should not include a hard constraint making the cone enclosing all the data, but should allow the extracted endmembers to be slightly inside the convex cone so as to be better representatives of each endmember class in when significant intrinsic variability is expected. With all these observations, we have proposed an algorithm to blindly unmix hyperspectral data in the presence of variability in difficult scenarios (correlated endmembers, shadow effects) and have shown its efficacy on a real and simulated dataset. Theoretical open questions include a rigorous study of the robustness of the volume regularization to noise and intrinsic variability of the endmembers. On the algorithmic part, future work will include a scheme to automatically estimate the regularization parameters of the algorithm.

\vspace{-0.3cm}
\bibliographystyle{ieeetr}
\bibliography{refs}
\end{document}